\documentclass[twocolumn]{revtex4}
\usepackage[dvipdfmx]{graphicx}
\usepackage{amsmath,amsbsy,amssymb}
\usepackage{bm}
\usepackage{mathrsfs,color,ulem}
\allowdisplaybreaks[1]

\newcommand{\diff}{\mathrm{d}}

\newcommand{\imu}{\mathrm{i}}
\newcommand{\epn}{\mathrm{e}}
\newcommand{\sgn}{\mathrm{sgn}\,}

\newcommand{\la}{\langle}
\newcommand{\ra}{\rangle}

\newcommand{\sg}{\sigma}
\newcommand{\gm}{\gamma}

\usepackage{color}

\begin{document}

\title{
Nature of superconducting fluctuation in photo-excited systems
}

\author{Ryuta Iwazaki$^1$, Naoto Tsuji$^2$ and Shintaro Hoshino$^1$}

\affiliation{
$^{1}$Department of Physics, Saitama University, Shimo-Okubo, Saitama 338-8570, Japan
\\
$^{2}$RIKEN Center for Emergent Matter Science (CEMS), Wako, Saitama 351-0198, Japan
}

\date{\today}

\begin{abstract}
The photo-excited state associated with superconducting fluctuation above the superconducting critical temperature $T_c$ is studied based on the time-dependent Ginzburg-Laundau approach.
The excited state is created by an electric-field pulse and is probed by a weak secondary external field, which is treated by the linear response theory mimicking pump-probe spectroscopy experiments.
The behavior is basically controlled by two relaxation rates: one is $\gm_1$ proportional to the temperature measured from the critical point $T - T_c$ and the other is $\gm_2$ proportional to the excitation intensity of the pump pulse.
The excited state approaches the equilibrium state exponentially in a long time $t \gg \gm_1^{-1}$, while in the intermediate time domain we find a power-law or logarithmic decay with different exponents for $t\ll \gm_2^{-1}$ and $\gm_2^{-1} \ll t \ll \gm_1^{-1}$, even though the system is located away from the critical point.
This is interpreted as the critical point in equilibrium being extended to a finite region in the excited situation.
The parameter dependences on both the pump and probe currents are also systematically studied in all dimensions.
\end{abstract}

\maketitle

\section{Introduction} \label{sec_intro}

Externally excited systems can show various exotic states that are not realizable as equilibrium states but nevertheless survive for a long time in non-equilibrium.
Such a possibility has been intensively studied by pump-probe experiments using a laser pulse, and the intriguing photo-excitated states have been created and measured \cite{Fausti11, Wang13, Stojchevska14, Zhang16, Iwai03, Perfetti06, Okamoto07, Giannetti16}.
For superconducting materials in strongly correlated systems, recent experiments have shown gigantic enhancement of superconductivity in a photo-induced state \cite{Fausti11, Kaiser14, Nicoletti14, Mitrano16, Hu14, Mankowsky14, Cantaluppi18, Giusti19}, which has considerable attention in the physics community.
While the change in the electron-phonon coupling through the phonon drive has been studied as the origin of the photo-induced superconducting state \cite{Knap16, Sentef16, Okamoto16, Kim16, Murakami17, Kennes17NatPhys, Nava18, Denny15, Sentef17, Babadi17, Mazza17}, photons also pump the electrons directly.
Some of the theoretical studies have proposed scenarios based on photo-excited electron systems \cite{Kennes17, Chiriaco18}.

The superconductivity-like behavior above the transition temperature $T_c$ is reminiscent of paraconductivity originating from the fluctuation effects: the electron conduction near the superconducting transition point becomes much higher than in the normal state.
Usually, the superconducting fluctuation is so small in three dimensional systems that it has been mainly discussed in low-dimensional systems \cite{Larkin_book}.
With excitations, however, the fluctuation effects may change.
Here, we study such a photo-induced phenomenon associated with the superconducting fluctuation based on the time-dependent (TD) Ginzburg-Landau (GL) approach.

The GL theory has been successfully applied to
unconventional superconductors \cite{Sigrist91}.
In the strongly correlated systems, since the coherence length can be an order of a lattice constant, the application of GL analysis may not be justified but still well describes their qualitative behaviors.
This indicates that extrapolation from the long wavelength (coherence length) limit works well for existing superconductors.
As for the time-dependent case, it is argued that the TDGL equation is not microscopically justified for gapped superconductors, because the relaxation time of quasiparticles can become longer than the time scale of the order parameter variation \cite{Gorkov68}.
On the other hand, once the TDGL equation is empirically introduced as a phenomenological theory, it has been recognized as a powerful tool for time-dependent phenomena such as vortex dynamics and fluctuation paraconductivity/diamagnetism \cite{Larkin_book, Kopnin_book}.
Furthermore, specifically for the paraconductivity defined in $T>T_c$, there is no gap in the excitation spectrum and the TDGL approach can be justified.

The TDGL theory has been applied to the pump-probe spectroscopy below the transition temperature \cite{Kennes17}.
With a fast pump process in real time, a broad range of frequency components are excited.
The TDGL equation can account for the dynamics of small-frequency components,
which should dominate the long-time behavior.
In this paper, we consider the superconducting fluctuation effects in the photo-excited state using the TDGL equation above the transition temperature, and the behaviors in the whole time scale can be uniquely determined.
The quench dynamics of superconducting fluctuations with a sudden change of an attractive interaction has been discussed in Ref.~\onlinecite{Lemonik18}.
We here consider purely electronic effects in photo-excited systems.
The original paper by Schmid has derived the fluctuation-induced current for arbitrary strength of electric fields \cite{Schmid69}.
Also, recently the non-linear response is considered in relation to the nonreciprocal transport phenomena in noncentrosymmetric systems \cite{Wakatsuki17, Wakatsuki18, Hoshino18}.
Thus the TDGL theory has the ability to tackle the problem beyond the usual linear-response regime.

When we apply the strong electric-field pulse,
there are two characteristic inverse-time scales $\gm_1$ and $\gm_2$.
$\gm_1$ originates from the reduced temperature measured from the critical temperature $T_c$, and determines a time constant for the relaxation into the equilibrium state.
The other one $\gm_2$ comes from the strength of the electric-field pulse.
We show that for $t\to \infty$, the system relaxes to the equilibrium state with the time constant $\gm_1^{-1}$, but
in the intermediate time domain, the power-law or logarithmically decaying regimes are found.
The exponents are different between the time regions $t\ll\gm_2^{-1}$ and $t \gg \gm_2^{-1}$.
As for the conductivity calculated from the probe current, the enhancement by the pulse is not observed.
On the other hand, the anisotropy shows a nontrivial behavior: although the system is spatially anisotropic due to the photo-induced current,
the excited current quickly damps compared to the superconducting fluctuation.
Hence we find an isotropic but still non-equilibrium state, which is confirmed through the properties of probe currents.
We also perform a systematic study on the parameter dependences on the pump and probe electric currents in all dimensions.

This paper is organized as follows.
Section~\ref{sec_procedure} provides the theoretical formalism for pump-probe procedure based on the TDGL approach.
The numerical results are given in Sec.~\ref{sec_result_single} and Sec.~\ref{sec_result_double} where different pump procedures are taken.
We summarize these results in Sec.~\ref{sec_summ}.
Appendices~\ref{app_dimless} and \ref{app_num} are devoted to the detailed explanation for our numerical calculation, and Appendix~\ref{app_eq} summarizes the equilibrium property realized for $t<0$ (before the pump) and $t\rightarrow \infty$.
The additional data for three and one dimensions are also shown in Appendix~\ref{sec_app_add_fig}.

\section{Pump-probe procedure in TDGL equation} \label{sec_procedure}

\subsection{General Formulation}

\begin{figure*}[t]
	\centering
    \includegraphics[width = 160mm]{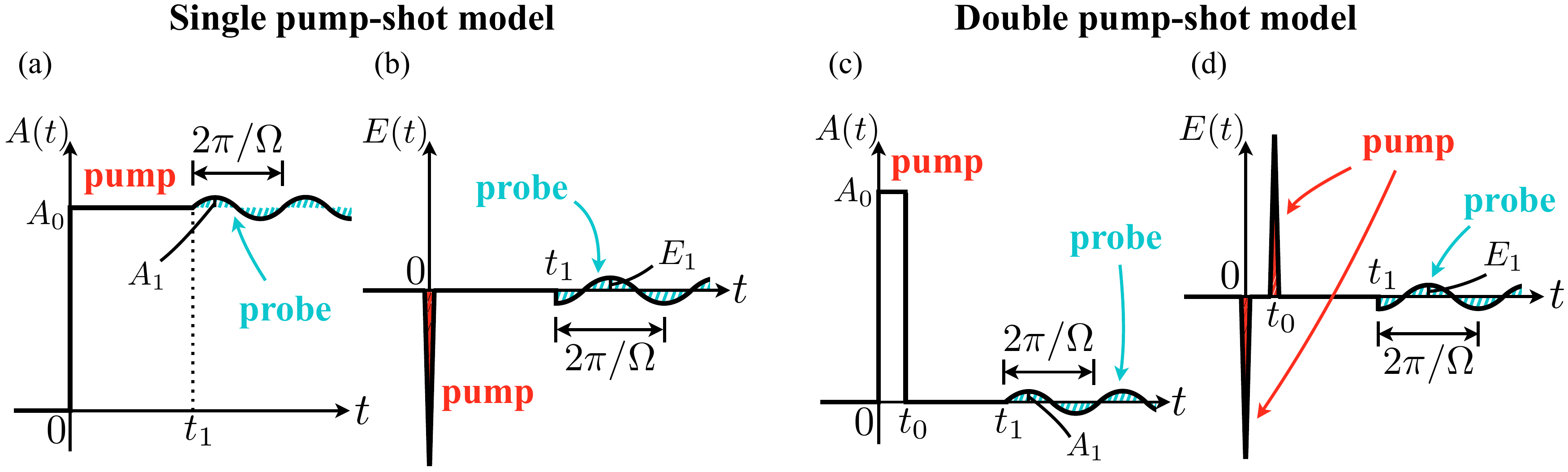}
	\caption{
    Sketches for the time dependences of the (a) vector potential and (b) electric field in the pump-probe process considered in the single pump-shot model.
    (c,d) The same plots
	for the double pump-shot model.
    }
	\label{fig:pump-probe}
\end{figure*}

Let us begin with the TDGL equation with a random force \cite{Schmid69}
\begin{align}
&\Gamma \frac{\partial}{\partial t} \psi(\bm r, t)
=- \frac{\delta \mathcal F[\psi,\bm A]}{\delta \psi^*(\bm r,t)}
 + f(\bm r, t),
\\
&\mathcal F[\psi,\bm A] = \hspace{-1mm} \int \hspace{-1mm} \diff \bm r
\psi^* (\bm r, t)\left(
a
-\frac{[\bm{\nabla} - 2\imu e\bm{A}(\bm r, t)]^2}{4m}
\right)
\psi(\bm r, t)
, \\
&\bm{j}(\bm{r}, t)
=
-\frac{\delta \mathcal F[\psi,\bm A]}{\delta \bm A(\bm r, t)}
,
\end{align}
where $\mathcal F$ and $\bm j$ are the GL free energy and superconducting current density, respectively, and $\Gamma$ is a time development factor.
 We have taken $\hbar = k_{\rm B} =1$ unless otherwise stated explicitly.
The classical complex field $\psi(\bm r, t)$ is the superconducting order parameter or macroscopic wave function.
We consider the spatially uniform electric field and the vector potential can be chosen as the one dependent only on time: $\bm{A}(\bm r, t) = \bm{A}(t)$.
We note that the fourth-order term $|\psi(\bm r, t)|^4$ is irrelevant here, since we consider the system above the transition temperature ($T > T_c$).
The photo-excitations in the superconducting state below $T_c$ has been discussed in Ref.~\onlinecite{Kennes17}.
The coefficient of the second-order term behaves as $a =a' (T - T_c)$ ($>0$) with an order-of-unity constant $a'$.
The random force $f(\bm r, t)$ is also introduced to represent the thermal fluctuation.
Performing the Fourier transformation, we get
\begin{align}
	\Gamma \frac{\partial}{\partial t} \psi_{\bm{q}}(t)
    =
    -\left(\frac{\left[\bm{q} - 2e\bm{A}(t)\right]^2}{4m} + a\right)\psi_{\bm{q}}(t) + f_{\bm{q}}(t).
\end{align}
The solution for the macroscopic wave function is then written as
\begin{align}
	&\psi_{\bm{q}}(t)
	=
	\nonumber
	\\
	&\int_{-\infty}^t \frac{\diff t'}{\Gamma} f_{\bm q}(t')
	\exp\left[
	- \int_{t'}^t \frac{\diff t''}{\Gamma} \left(
	\frac{[\bm q- 2e\bm A(t'')]^2}{4m} + a
	\right)
	\right].
\end{align}
We assume that $f_{\bm{q}}(t)$ is spatially and temporally uncorrelated
such that the random average (written by a bracket symbol) is given by
\begin{align} \label{random_ave}
	\left< f_{\bm{q}}^{\ast}(t) f_{\bm{q'}}(t')\right>
	&=
	\frac{C}{V} \delta_{\bm q\bm q'}\delta(t - t'),
\end{align}
where $V$ is a system volume.
The constant $C$ is determined so that the random average is identical to the thermal average \cite{Schmid69}, which results in $C=2\Gamma VT$.

The most fundamental physical quantity in our system is the superconducting fluctuation $\la |\psi|^2\ra$.
With the random force, the effect of the superconducting fluctuation can be taken into account and the quantity $\la |\psi_{\bm q}(t)|^2 \ra$ becomes finite.
Since the quantity with the wavevector $\bm q$ is gauge-dependent, we consider the real-space quantity
\begin{align}
\la |\psi(\bm r, t)|^2 \ra &= \frac{1}{V} \sum_{\bm q} \la |\psi_{\bm q}(t)|^2 \ra ,
\end{align}
which is now a gauge-invariant quantity, and is also spatially uniform in our present setup.
In the analysis, we separate the contributions from the equilibrium state and the non-equilibrium one, and define the deviation by
\begin{align}
	\delta \la |\psi_{\bm q}(t)|^2 \ra
	&=
	\la |\psi_{\bm q}(t)|^2 \ra -
	\la |\psi_{\bm q}(t)|^2 \ra_\mathrm{eq},
\end{align}
where $\la \cdots \ra_{\rm eq}$ stands for the equilibrium value.
In the following, we use $\delta O$ as the deviation of the quantity $O$ from that in equilibrium state.

We separate the vector potential into two parts as $\bm A= \bm A_0 + \bm A_1$ where $\bm A_0$ and $\bm A_1$ represent pump and probe lights, respectively.
The vector potential $\bm A_1$ for the probe is treated by the linear-response theory.
The fluctuation contribution to the electric current density is given by
\begin{align}
	\bm j (t)
	&=\sum_{\bm q} \frac{e}{mV} [\bm q - 2e\bm A(t)] \left< |\psi_{\bm q}(t)|^2 \right>
=\bm{j}_0(t) + \bm{j}_{1}(t),
\label{def_current}
\end{align}
where
\begin{align}
\bm j_0(t) &= \frac{e}{mV} \sum_{\bm q} [\bm q - 2e\bm A_0(t)] \left< |\psi_{\bm q}(t)|^2 \right>_0,
\label{eq:j0}
\\
\bm j_1(t) &= \frac{e}{mV} \sum_{\bm q} [\bm q - 2e\bm A_0(t)] \left< |\psi_{\bm q}(t)|^2 \right>_1
\nonumber \\
&\ \ \ - \frac{2e^2}{mV}\sum_{\bm q} \left< |\psi_{\bm q}(t)|^2 \right>_0 \bm A_1(t).
\label{eq:j1}
\end{align}
Here $\bm j_0$ is a photo-excited current density and $\bm j_1$ is a small current induced by the probe light.
The subscripts 0 and 1 represent the perturbation order of $\bm A_1$.
If we consider the equilibrium case, we reproduce the standard paraconductivity formula \cite{Larkin_book, Schmid69}, which are summarized in Appendix~\ref{app_eq} based on our formalism.

We can also consider the time-dependent GL free energy $\mathcal F$, which is modified from the equilibrium by the photo-excitation.
After taking the random average, we get
\begin{align} \label{def_free_energy}
	\left< \mathcal{F}(\bm{r}, t) \right>
	=&
	\sum_{\bm{q}}\left( a + \frac{\left[\bm{q} - 2e\bm{A}(t)\right]^2}{4m} \right) \la |\psi_{\bm{q}} (t)|^2 \ra ,
	\\
	\equiv&
\mathcal{F}_{\mathrm{fluct}}(t)
	+
\mathcal{F}_{\mathrm{kin}}(t). \label{def_free_energy_component}
\end{align}
The first term $\mathcal{F}_{\mathrm{fluct}}(t) = V a \la |\psi(\bm r, t)|^2 \ra$ represents the fluctuation contribution to the free energy, and the second term represents the kinetic energy of Cooper pairs.
In order to consider the anisotropy in the kinetic energy, we divide the free energy as $\mathcal F_{\rm kin}=\sum_\mu \mathcal F_{\rm kin}^{(\mu)}$ where
	\begin{align}
	\mathcal{F}_{\rm kin}^{(\mu)}(t)
	=&
	\sum_{\bm{q}} \frac{[q_\mu - 2eA_\mu (t)]^2}{4m} \left< |\psi_{\bm{q}} (t)|^2 \right>.
	\label{eq:fkin_decomp}
	\end{align}
In the next two subsections, we specify the shape of the pump/probe lights more concretely, and derive the detailed expressions for the above physical quantities.

\vspace{5mm}

\subsection{Single pump-shot model} \label{sec_proc_single}

As the simplest form of the pump process, we consider the electric field with a single delta function [single pump-shot model, see Figs.~\ref{fig:pump-probe}(a) and (b)].
The time dependence of the vector potential is
\begin{align}
	\bm A (t) = \bm A_0 \theta (t) + \bm A_1 \theta (t-t_1) \sin \Omega (t-t_1),
\end{align}
where $\theta (t)$ is Heaviside's step function.
While the realistic setup in the experiments involves an oscillating Gaussian shape, our choice of electric fields makes the equations simple enough to evaluate decaying properties precisely in numerical calculations.
More specifically, the integration with respect to time is evaluated analytically, and hence we can focus on the implementation of $\bm q$-integrals.
With this setup, the superconducting fluctuation is explicitly written as
\begin{widetext}
	\begin{align}
		&\delta \left< |\psi_{\bm q}(t)|^2 \right> =
		\frac{C}{V\Gamma^2 g(\bm{q})^{-1}}
		\epn^{
		- t g(\bm{q} - 2e\bm{A}_0)^{-1}
		}
		\left(
		1 - \frac{2e(\bm q- 2e\bm A_0)\cdot \bm A_1\theta (t-t_1)}{m\Gamma \Omega} [\cos \Omega (t - t_1) - 1]
		\right)
		\nonumber \\
		&\hspace{21mm}-
		\frac{C}{V\Gamma^2 g(\bm{q} - 2e\bm{A}_0)^{-1}}
		\epn^{
		- t g(\bm{q} - 2e\bm{A}_0)^{-1}
		}
		\left(
		1 - \frac{2e(\bm q- 2e\bm A_0)\cdot \bm A_1\theta (t-t_1)}{m\Gamma \Omega} [\cos \Omega (t - t_1) - 1]
		\right)
		\nonumber \\
		&\hspace{21mm}-
		\frac{C}{V\Gamma^2}
		\cdot
		\frac{2e(\bm q- 2e\bm A_0)\cdot \bm A_1\theta (t-t_1)}{m\Gamma\Omega} \cdot
		\frac{g(\bm q- 2e\bm A_0)^{-1}}{g(\bm q- 2e\bm A_0)^{-2}+\Omega^2} \epn^{-(t-t_1) g(\bm q- 2e\bm A_0)^{-1}} + \mathcal{O}(\bm A_1^2),
		\label{single_fluc}
		\\
		&\left< |\psi_{\bm q}(t)|^2\right>_\mathrm{eq} =
		\frac{C}{V\Gamma^2 g(\bm q-2e\bm A_0)^{-1}}
		\left(
		1 - \frac{2e(\bm q-2e\bm A_0)\cdot \bm A_1\theta (t-t_1)}{m\Gamma\Omega}
		\cos \Omega(t-t_1)
		\right)
		\nonumber
		\\
		&+
		\frac{C}{V\Gamma^2}
		\cdot
		\frac{2e(\bm q-2e\bm A_0)\cdot \bm A_1\theta (t-t_1)}{m\Gamma\Omega}
		\left[
		\frac{g(\bm q-2e\bm A_0)^{-1}}{g(\bm q-2e\bm A_0)^{-2}+\Omega^2} \cos\Omega(t-t_1)
		+ \frac{\Omega}{g(\bm q-2e\bm A_0)^{-2}+\Omega^2} \sin\Omega(t-t_1)
		\right] + \mathcal{O}(\bm A_1^2)
, \label{psi_eq_probe}
	\end{align}
\end{widetext}
for $t>0$.
We have defined the $\bm q$-dependent function by $g(\bm{q}) = \frac{\Gamma}{2}(\frac{\bm{q}^2}{4m} + a)^{-1}$ to make the notation simple.
There is no exponentially time-dependent factor in $\left< |\psi_{\bm{q}}(t)|^2 \right>_{\mathrm{eq}}$, and hence it represents the contribution from equilibrium in the limit $t\to \infty$.
Taking the $\bm q$-summations with some form factors, we can get physical quantities such as the order parameter and electric current, which are all gauge-invariant quantities.
We note that even after the pump process finishes, there remains finite vector potential as shown in Fig.~\ref{fig:pump-probe}(a).
However, this is not a physical degrees of freedom, and it can be eliminated by a constant shift in $\bm q$-summation.
We also note that the expression includes exponential functions and the integration range should be carefully chosen in the numerical calculation.
The technical details are summarized in Appendices~\ref{app_dimless} and \ref{app_num}.
With these techniques, we can have highly accurate numerical results which enable us to access the critical exponents of the decaying functions as shown later in Sec.~\ref{sec_result_single}.

Although in general it is hard to derive the analytical expression, it is possible if we focus on the large $t$ limit.
In this case the $\bm{q}$-integrals are reduced to simple Gaussian integrals, and the superconducting fluctuation, electric current density and kinetic component of the GL free energy are explicitly obtained as
	\begin{align}
		&\delta \la |\psi(\bm{r}, t)|^2 \ra \xrightarrow[]{t \to \infty} \nonumber \\
		&\hspace{4mm} \frac{C}{2\Gamma V} \left( \frac{m\Gamma}{2\pi}\right)^{d/2} \left( \frac{m}{e^2 \bm A_0^2 + ma} - \frac{1}{a} \right) t^{-d/2}
		\epn^{-\frac{2a}{\Gamma}t},
		\label{eq:analytic_single_shot_fluc}
		\\
		&\bm j_{0}(t) \xrightarrow[]{t \to \infty} \nonumber \\
		&\hspace{8mm} -\frac{C e^2 \bm A_0}{\Gamma V} \left( \frac{m\Gamma}{2\pi}\right)^{d/2}
		\frac{1}{e^2\bm{A}_0^2 + ma} \, t^{-d/2}
		\epn^{-\frac{2a}{\Gamma}t},
		\label{eq:analytic_single_shot_curr}
		\\[2mm]
		&\delta \mathcal{F}_{\mathrm{kin}}(t) \xrightarrow[]{t \to \infty} \nonumber \\
		&\hspace{5mm} \frac{dC}{8} \left( \frac{m\Gamma}{2\pi}\right)^{d/2} \left( \frac{m}{e^2 \bm A_0^2 + ma} - \frac{1}{a} \right) t^{-\frac{d+2}{2}}
		\epn^{-\frac{2a}{\Gamma}t},
		\label{eq:analytic_single_shot_kin}
	\end{align}
for the $d$-dimensional system.
Namely, the time constant $\Gamma/2a$ (which is identical to $\gm_1^{-1}$ defined below in Eq.~\eqref{def_gm1}), which originates from the temperature measured from the transition point, gives a characteristic time scale when approaching the equilibrium state.
However, the richer behaviors can be seen at shorter time range, as discussed in Sec.~\ref{sec_result_single}.

\vspace{5mm}

\subsection{Double pump-shot model} \label{sec_proc_double}

Here we consider a slightly more realistic situation.
Since the electric field is applied with an oscillating manner in the experimental setup,
it is natural if the sum rule $\int_0^{t_0} \bm{E}(t) \diff t = \bm{0}$ is satisfied.
With this constraint, the simplest form of the electric field is the combination of the two delta functions with different signs [see Figs.~\ref{fig:pump-probe}(c) and (d)].
The time dependence of the vector potential is given by
\begin{align}
	\bm{A}(t)
	&= \bm{A}_0 \theta(t) \theta(t_0 - t)
	+ \bm{A}_1 \theta (t-t_1) \sin \Omega (t - t_1).
\end{align}
Namely, the pump process appears for $0<t<t_0$ and the probe process for $t_1 < t$.
In this case, the deviation of the superconducting fluctuation from the equilibrium state for $t>t_0$ is written as
\begin{widetext}
	\begin{align}
		&\delta \left< |\psi_{\bm q}(t)|^2\right> =
		\frac{C}{V\Gamma^2 g(\bm q)^{-1}}
		\epn^{
		- (t-t_0) g(\bm q)^{-1}
		- t_0 g(\bm q-2e\bm A_0)^{-1}
		}
		\left(
		1 - \frac{2e\bm q\cdot \bm A_1\theta(t-t_1)}{m\Gamma\Omega}\left[\cos \Omega(t-t_1) - 1 \right]
		\right)
		\nonumber \\
		&\hspace{0mm}+
		\frac{C}{V\Gamma^2 g(\bm q-2e\bm A_0)^{-1}}
		\left[
		 1
		- \epn^{ - t_0 g(\bm q-2e\bm A_0)^{-1} }
		\right]
		\epn^{
		- (t-t_0) g(\bm q)^{-1}
		}
		\left(
		1 - \frac{2e\bm q\cdot \bm A_1\theta(t-t_1)}{m\Gamma\Omega}\left[\cos \Omega(t-t_1) - 1 \right]
		\right)
		\nonumber \\
		&\hspace{0mm}+
		\frac{C}{V\Gamma^2 g(\bm q)^{-1}}
		\left[
		\epn^{
		- (t-t_1) g(\bm q)^{-1}
		}
		-
		\epn^{
		- (t-t_0) g(\bm q)^{-1}
		}
		\right]
		\left(
		1 - \frac{2e\bm q\cdot \bm A_1\theta(t-t_1)}{m\Gamma\Omega}\left[\cos \Omega(t-t_1) - 1 \right]
		\right)
		\nonumber \\
		&\hspace{0mm}-
		\frac{C}{V\Gamma^2 g(\bm{q})^{-1}}
		\epn^{-(t - t_1) g(\bm{q})^{-1}}
		\left(
		1 - \frac{2e\bm{q} \cdot \bm{A}_1\theta(t-t_1)}{m\Gamma\Omega} \cos\Omega(t-t_1)
		\right)
		-
		\frac{C}{V\Gamma^2}
		\cdot
		\frac{2e\bm q\cdot \bm A_1\theta(t-t_1)}{m\Gamma\Omega}
		\cdot
		\frac{\epn^{-(t-t_1) g(\bm q)^{-1}} g(\bm q)^{-1}}{g(\bm q)^{-2}+\Omega^2},
		\label{double_fluc}
	\end{align}
\end{widetext}
where we have kept the terms up to $\mathcal{O}(\bm A_1)$ contributions.
The order parameter in the equilibrium state is the same as Eq.~\eqref{psi_eq_probe} with the replacement $\bm{q} - 2e\bm{A}_0 \to \bm{q}$.
For the time range $0<t<t_0$, the system is essentially the same as the single pump-shot model and new expressions are not necessary.

\section{Numerical Results for single pump-shot model} \label{sec_result_single}
\begin{figure*}[t]
	\centering
	\includegraphics[width=160mm]{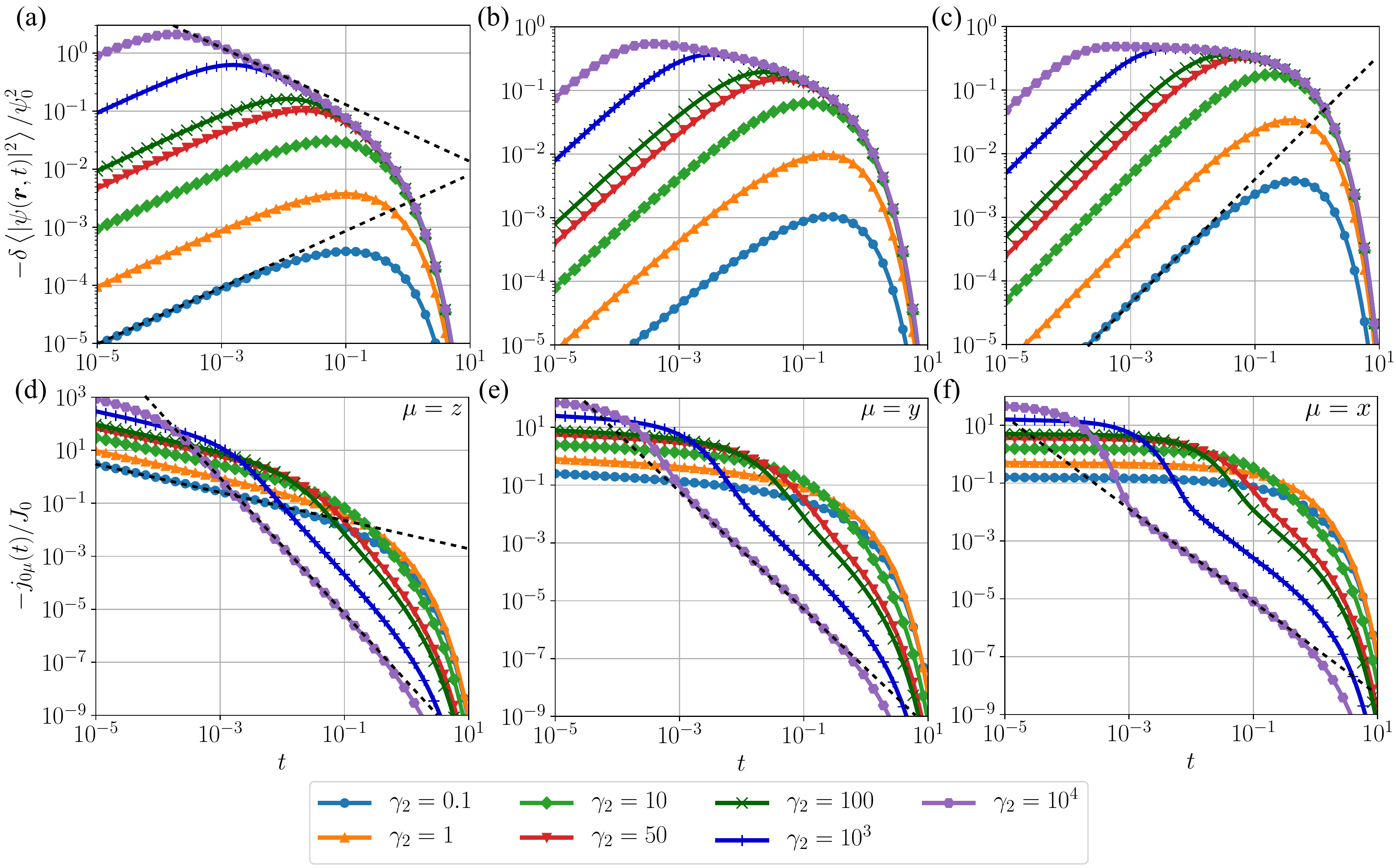}
	\caption{
		Time evolution of the (a--c) superconducting fluctuation and the (d--f) excited current density in the single pump-shot model with a logarithmic scale, which are measured as deviations from the equilibrium component.
		The quantities are plotted for (a,d) three-, (b,e) two- and (c,f) one-dimensional systems.
		The dashed lines shown in (a,c--f) are power functions shown in Eqs.~\eqref{eq:exponent_fluc_1}, \eqref{eq:exponent_curr_1}, \eqref{eq:exponent_fluc_2} and \eqref{eq:exponent_curr_2}.
		$\mu$ is the direction where the pump pulse is shotted, $\mu = z$ in the (d) three-, $\mu = y$ in the (e) two- and $\mu = x$ in the (f) one-dimensional system.
		The normalized constants are defined by $\psi_0^2 = \frac{C}{\gm_1 \Gamma^2 V} \left( 2m\gm_1 \Gamma \right)^{d/2}$ and $J_0 = \frac{eC}{m\gm_1 \Gamma^2 V} \left( 2m\gm_1 \Gamma \right)^{\frac{d+1}{2}}$.
	}
	\label{fig:single_fluc_curr}
\end{figure*}
Here we show the physical quantities in the photo-excited states of the single-pump shot model formulated in Sec.~\ref{sec_proc_single}.
We concentrate on the photo-excited states in this section, and the properties of probing process are discussed in the more realistic double-pump shot model in the next section.
For the single-pump shot model, the behaviors are controlled by the following two parameters
\begin{align} \label{def_gm1}
	\gm_1 =& \frac{2a}{\Gamma},
	\\
	\gm_2 =& \frac{2e^2 \bm A_0^2}{m\Gamma},
	\label{def_gm2}
\end{align}
which have the dimension of the inverse of time.
$\gm_1$ represents the distance from the critical point and $\gm_2$ shows the intensity of the pump light.
We take $\gm_1=1$ as a unit of energy, and the controlling parameter is then $\gm_2$.

First, we consider the deviation $\delta \la |\psi(\bm r, t)|^2 \ra$ of the superconducting fluctuation from equilibrium defined in Eq.~\eqref{single_fluc}, which is shown in Figs.~\ref{fig:single_fluc_curr}(a), (b) and (c).
These figures are plotted with a logarithmic scale to visualize the functional form of the decaying functions of the physical quantities.
After the pump at $t=0$, the superconducting fluctuation decreases from the equilibrium value, i.e., $\delta \la |\psi(\bm r, t)|^2 \ra < 0$.
The decrease stops at $t\sim \gm_2^{-1}$, where the fluctuation becomes minimum, and then it increases toward the equilibrium value.
It is notable that in the whole time range the photo-excitation gives negative contribution to the superconducting fluctuation, which is in contrast to a naive expectation that the photo-excitation may increase the superconducting fluctuation.
The photo-induced electric current is also shown in Figs.~\ref{fig:single_fluc_curr}(d--f), which is a decreasing function with respect to time.
Namely, the effect of photo-excitation appears maximally around $t=0$, which is different from the behaviors of the fluctuation shown in Figs.~\ref{fig:single_fluc_curr}(a--c).

For a sufficiently long time regime, all the quantities in Fig.~\ref{fig:single_fluc_curr} decay with an exponential form with the damping constant $\gm_1$ ($=1$).
Indeed, this behavior is consistent with the analytical asymptotic form given in Eqs.~\eqref{eq:analytic_single_shot_fluc} and \eqref{eq:analytic_single_shot_curr}.
In the shorter time regime, on the other hand, we find the power-law or logarithmic behaviors for both the superconducting fluctuation and electric current in all dimensions.
Usually the power-law behaviors are observed in systems located at the critical point \cite{Hohenberg77}.
In contrast, the present case is away from the transition point ($T>T_c$) but we still observe the power-law decay.
In large $\gm_2$ cases such as $\gm_2=50$, $100$, $10^3$ and $10^4$, the power-law regime is separated into two parts at $t\sim \gm_2^{-1}$.
We can see from Fig.~\ref{fig:single_fluc_curr} that if we take $\gm_2 < \gm_1$, one of the power-law or logarithmic regimes vanishes.
In the former studies in non-equilibrium states, the power-law behavior in superconductors has also been found in Refs.~\onlinecite{Volkov73} and \onlinecite{Barankov06}.

\begin{figure}[t]
	\centering
	\includegraphics[width=85mm]{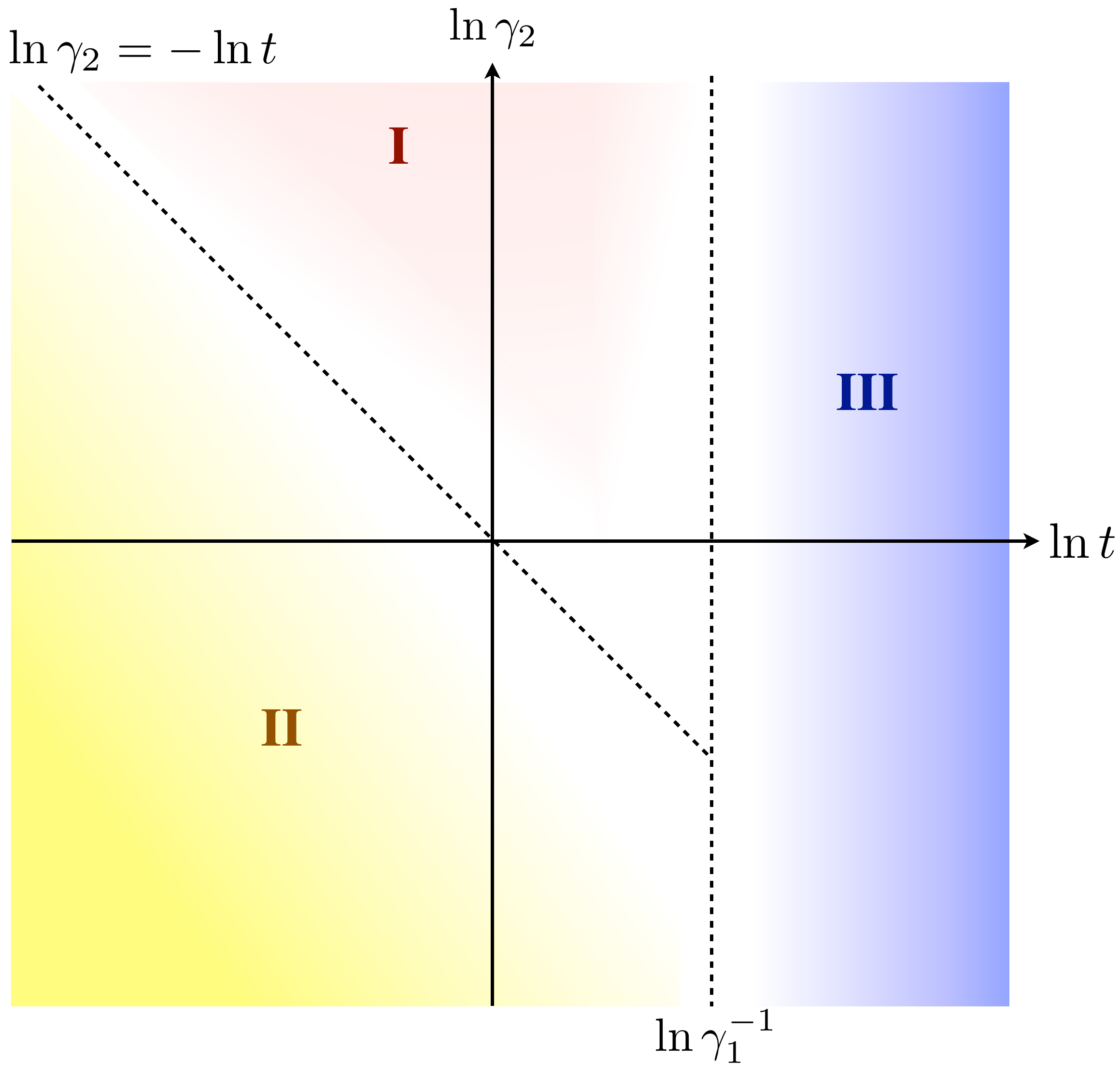}
	\caption{
		A sketch of the phase diagram in the plane of the logarithmic time $\ln t$ and the logarithmic excitation strength $\ln \gm_2$.
		The dashed lines show crossovers.
	}
	\label{fig:phase}
\end{figure}

The above exponential and power-law/logarithmic behaviors are summarized in Fig.~\ref{fig:phase}, which is one of the central results of this paper.
The phase diagram in the plane of $t$ and $\gm_2$ is categorized into the three regimes:
the exponentially decaying regime ($t>\gm_1^{-1}$, III), the power-law or logarithmic decaying regime ($\gm_{2}^{-1}<t<\gm_1^{-1}$, I), and another power-law regime ($t<\gm_2^{-1}$, II).
The pump-pulse intensity $\gm_2$ characterizes the power-law/logarithmic damping behavior, while the system temperature $\gm_1 \propto (T-T_c)$ characterizes exponentially decaying behavior.
The line that separates the regimes I and II terminates at the point where it crosses with the vertical $t=\gm_1^{-1}$ line.

The power-law/logarithmic behavior is intuitively interpreted as follows.
When the system is very close to the superconducting transition point, the value of $\gm_1^{-1} \sim a^{-1}$ goes to infinity and then only the regimes I and II are left.
Hence these power-law/logarithmic decay originates from a characteristic behavior at the critical point.
For a sufficiently weak pump, the regime II survives for a long time, whereas the regime I is dominant for a strong pump with large $\gm_2 \propto \bm A_0^2$.
Away from the critical point ($a>0$), the power-law/logarithmic regimes end at the time scale of $\gm_1^{-1}$ and crossovers into the exponentially decaying regime.
This indicates that the critical point is effectively expanded into a finite regime by the pump shot and it shrinks with evolving time to approach the equilibrium state.
We note that the fluctuation itself is suppressed as shown in Figs.~\ref{fig:single_fluc_curr}(a) and (c) and Eq.~\eqref{single_fluc}, although it is divergent at the critical point in equilibrium.
Hence the behavior with the power-law/logarithmic decay and with finite fluctuation amplitude is characteristic for an excited state in non-equilibrium.

We also comment on the region in which the GL approach is justified.
The GL theory can account for the long wavelength and slow dynamics, and the corresponding characteristic length and time are given by the coherence length $\xi \sim \hbar / \sqrt{ma}$ and the correlation time $t_c \sim \Gamma / a \sim \hbar / a \sim \gm_1^{-1}$.
Hence the time range with $t\ll t_c \sim \gm_1^{-1}$ is not justified in a strict sense, but rather our results should be regarded as an extrapolation from the GL theory.

With the help of Fig.~\ref{fig:phase}, we can have analytic forms of the power-law and logarithmic functions.
For the regime I ($\gm_2^{-1} < t < \gm_1^{-1}$), we take the limit $\gm_2\to \infty$ and obtain the asymptotic behaviors of fluctuations and currents for each dimension as
	\begin{align}
		&\delta \la |\psi (\bm r,t)|^2 \ra
		\simeq
		\begin{cases}
			-\cfrac{4\pi mC}{V} \sqrt{\cfrac{2\pi m}{\Gamma}}\, \, t^{-1/2} &(d = 3),
			\\[10pt]
			\cfrac{2\pi m C}{\Gamma V} \left[ \gm + \ln \left( \cfrac{2a}{\Gamma}\, t\right) \right] &(d = 2),
			\\[10pt]
			-\cfrac{\pi C}{\Gamma V} \sqrt{\cfrac{m}{a}} &(d = 1),
		\end{cases}
		\label{eq:exponent_fluc_1}
		\\[10pt]
		&\bm j_{0}(t)
		\simeq
		\begin{cases}
			-\cfrac{2\pi m^2\Gamma C \bm A_0}{V e^2 \bm A_0^4} \sqrt{\cfrac{\pi m\Gamma}{2}}\, \, t^{-5/2} &(d = 3),
			\\[10pt]
			-\cfrac{\pi m^2 C\bm A_0}{V e^2 \bm A_0^4}\, \, t^{-2} &(d = 2),
			\\[10pt]
			-\cfrac{m C\bm A_0}{V e^2 \bm A_0^4} \sqrt{\cfrac{\pi m\Gamma}{2}}\, \, t^{-3/2} &(d = 1),
		\end{cases}
		\label{eq:exponent_curr_1}
	\end{align}
where $\gm$ is the Euler's gamma constant.
For the regime II ($t < \gm_2^{-1}$), on the other hand, the critical behaviors are seen in the weak excitation limit($\gm_2 \to 0$).
We then obtain the time dependences as
	\begin{align}
		&\delta \la |\psi (\bm r,t)|^2 \ra
		\simeq
		\begin{cases}
			-\cfrac{2e^2 \bm{A}_0^2 C}{3\pi \Gamma^2 V} \sqrt{\cfrac{2m}{\pi\Gamma}}\, \, t^{1/2} &(d = 3),
			\\[10pt]
			\cfrac{e^2 \bm{A}_0^2 C}{\pi \Gamma^2 V}\, \, t \left[ \gm + \ln \left( \cfrac{2a}{\Gamma}\, t\right) \right] &(d = 2),
			\\[10pt]
			-\cfrac{e^2 \bm{A}_0^2 C}{m\Gamma^2 V} \sqrt{\cfrac{m}{a}}\, \, t &(d = 1),
		\end{cases}
		\label{eq:exponent_fluc_2}
		\\[10pt]
		&\bm j_{0}(t)
		\simeq
		\begin{cases}
			-\cfrac{16\pi C e^2 \bm A_0}{3V} \sqrt{\cfrac{2\pi m}{\Gamma}}\, \, t^{-1/2} &(d = 3),
			\\[10pt]
			\cfrac{4\pi C e^2 \bm A_0}{\Gamma V} \left[ \gm + \ln \left( \cfrac{2a}{\Gamma}\, t\right) \right] &(d = 2),
			\\[10pt]
			-\cfrac{2\pi C e^2 \bm A_0}{V} \sqrt{\cfrac{2}{ma}} &(d = 1).
		\end{cases}
		\label{eq:exponent_curr_2}
	\end{align}
These time dependences are consistent with the numerical results shown in Fig.~\ref{fig:single_fluc_curr}.

\begin{figure}[t]
	\centering
	\includegraphics[width=85mm]{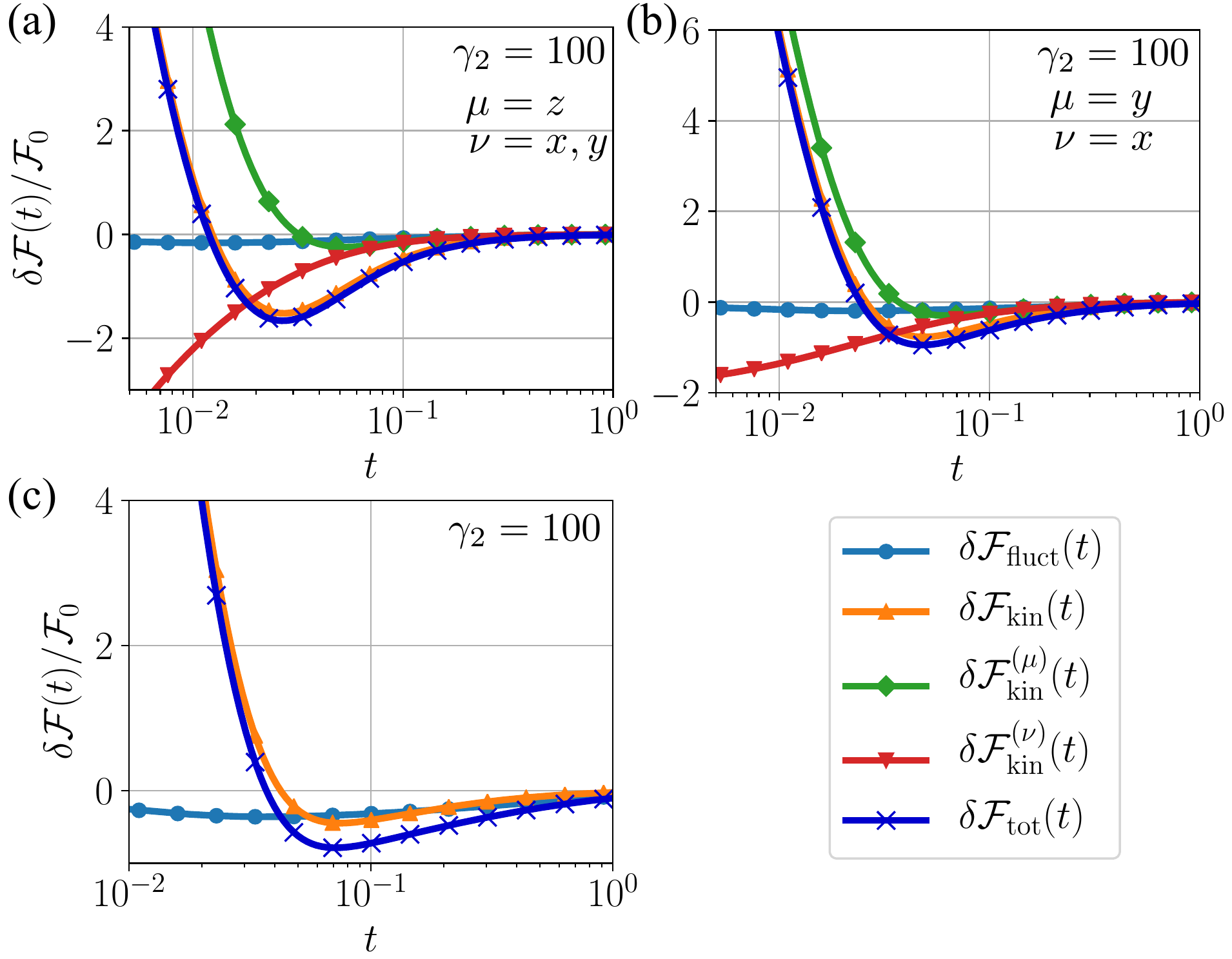}
	\caption{
		Time dependence of the GL free energy defined in Eq.~\eqref{def_free_energy} in the single pump-shot model($\delta \mathcal{F}_{\mathrm{tot}} = \delta \mathcal{F}_{\mathrm{fluct}} + \delta \mathcal{F}_{\mathrm{kin}}$) in the (a) three-, (b) two- and (c) one-dimensional systems and we have taken $\gm_2 = 100$.
		$\mu$ and $\nu$ mean the components of the kinetic GL free energy, where $\mu$ is the direction where the pump pulse is shotted and $\nu$ is the other direction,
		$\mu = z$ and $\nu = x, y$ in the (a) three dimensional system and $\mu = y$ and $\nu = x$ in the (b) two dimensional system.
		The normalized constant is defined by $\mathcal F_0 = \frac{C}{2\Gamma} \left( 2m\gm_1 \Gamma \right)^{d/2}$.
	}
	\label{fig:free_energy}
\end{figure}

Let us turn our attention to the photo-excitation effects on the free energy.
Figures~\ref{fig:free_energy}(a--c) show the time dependence of the total GL free energy $\delta \mathcal F(t)$ defined in Eq.~\eqref{def_free_energy} for the fixed $\bm A_0$.
The total free energy $\delta \mathcal F(t)$ substantially increases after the pump, whose main contribution comes from the kinetic energy.
We also plot the direction-dependent kinetic energy $\delta \mathcal F_{\rm kin}^{(\mu)}$ defined in Eq.~\eqref{eq:fkin_decomp} and the fluctuation contribution $\delta \mathcal F_{\rm fluct}$.
For the three dimensional case shown in Fig.~\ref{fig:free_energy}(a), the kinetic energy along $z$ direction is positive, while the other $x,y$-direction components are negative.
With evolving time, the kinetic energy decreases and the non-equilibrium component of the total free energy becomes negative.
This sign change is captured by analytic asymptotic functions for the kinetic energy, whose forms are
	\begin{align}
		&\delta \mathcal{F}_{\mathrm{kin}}(t)
		\sim
		\begin{cases}
			-\cfrac{mC}{4\pi} \sqrt{\cfrac{m\Gamma}{2\pi}}\, \, t^{-3/2} &(d = 3),
			\\[10pt]
			-\cfrac{m C}{4\pi}\, \, t^{-1} &(d = 2),
			\\[10pt]
			-\cfrac{C}{2\Gamma} \sqrt{\cfrac{m\Gamma}{2\pi}}\, \, t^{-1/2} &(d = 1),
		\end{cases}
		\label{eq:exponent_free_kin_1}
	\end{align}
	for $\gm_2^{-1} < t < \gm_1^{-1}$ (regime I), and
	\begin{align}
		&\delta \mathcal{F}_{\mathrm{kin}}(t)
		\sim
		\begin{cases}
			\cfrac{e^2\bm{A}_0^2 C}{6\pi \Gamma} \sqrt{\cfrac{2m\Gamma}{\pi}}\, \, t^{-1/2} &(d = 3),
			\\[10pt]
			-\cfrac{e^2\bm{A}_0^2 C}{2\pi} \left[ \gm + \ln \left( \cfrac{2a}{\Gamma}\, t\right) \right] &(d = 2),
			\\[10pt]
			\cfrac{e^2\bm{A}_0^2 C}{2m\Gamma} \sqrt{\cfrac{m}{a}} &(d = 1),
		\end{cases}
		\label{eq:exponent_free_kin_2}
	\end{align}
	for $t < \gm_2^{-1}$ (regime II).
The fluctuation contribution $\delta \mathcal F_{\mathrm{fluct}}(t)$ is basically the same quantity as the superconducting fluctuation given by $ \delta \la |\psi(\bm r,t)|^2 \ra$.
In Fig.~\ref{fig:free_energy}, however, it does not give a significant contribution to the free energy.

\section{Numerical results for double pump-shot model} \label{sec_result_double}

We consider the more realistic double pump-shot model shown in Figs.~\ref{fig:pump-probe}(c) and (d) as introduced in Sec.~\ref{sec_proc_double}.
We take $t_0=1$ as a unit of time, which is located at the end of the pump process [see Figs.~\ref{fig:pump-probe}(c) and (d)].
In this section, we discuss only the two-dimensional system because the behaviors are similar in the other dimensions.
The results for the one- or three-dimensional system are shown in Appendix~\ref{sec_app_add_fig} to make the data complete.

\subsection{Properties of excited state}

\begin{figure}[t]
	\centering
	\includegraphics[width=85mm]{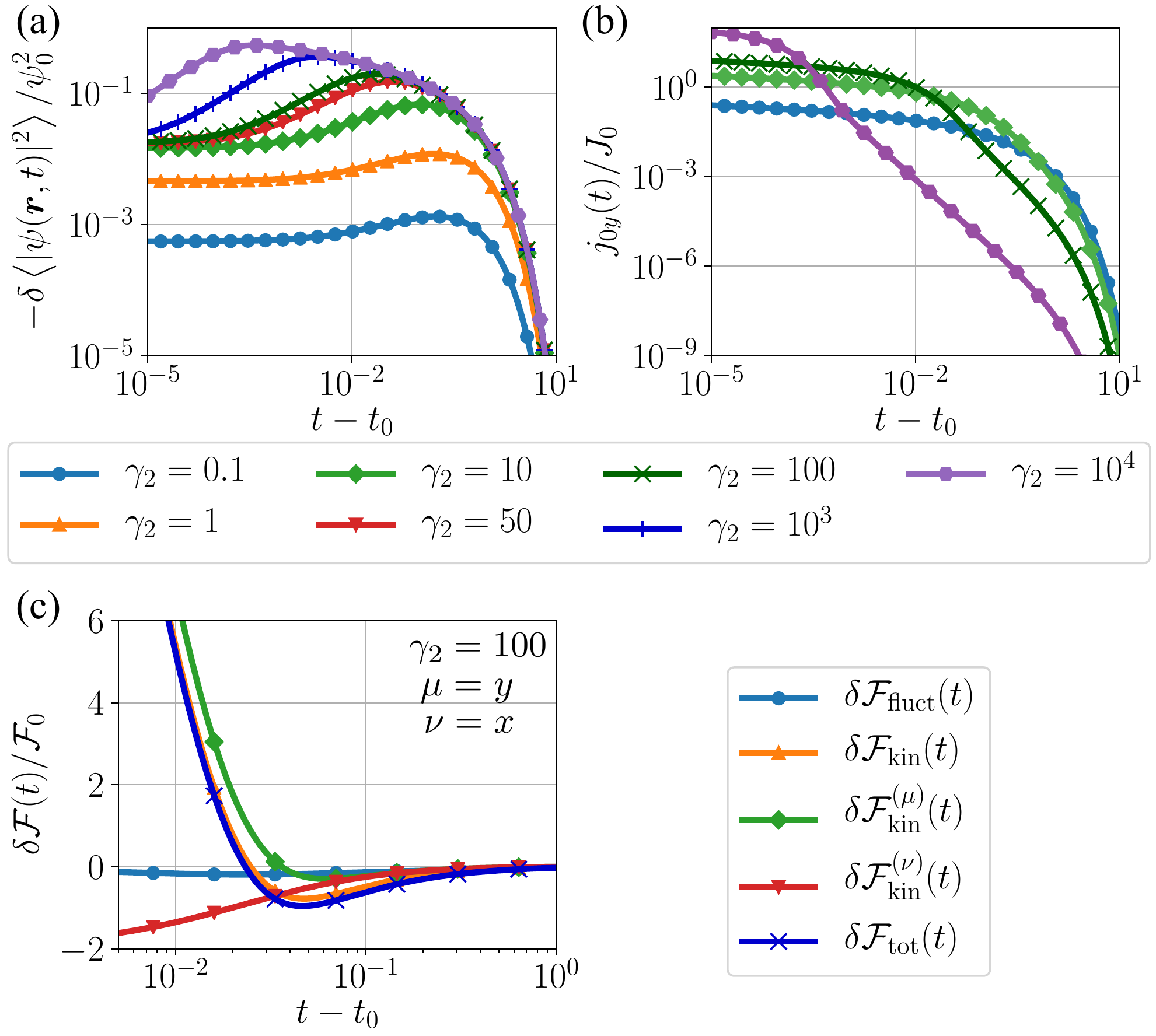}
	\caption{
		Time dependences of the (a) superconducting fluctuation, the (b) excited current density and the (c) GL free energy with the logarithmic time scale in the double pump-shot model.
		We have taken $t_0 = 1$, $\gm_1 = 1$ and (c) $\gm_2 = 100$ to compare with the results of the single pump-shot model in the two-dimensional system.
		$\mu$ and $\nu$ used in (c) have the same meanings as Fig.~\ref{fig:free_energy}(b).
		The normalized constants are defined by $\psi_0^2 = \frac{2mC}{\Gamma V}$, $J_0 = \frac{2eC}{\Gamma V} \sqrt{\frac{2m\Gamma}{t_0}}$ and $\mathcal{F}_0 = \frac{mC}{t_0}$.
	}
	\label{fig:3d_double_2nd_pump}
\end{figure}
We first summarize the results for the excited state here, and then discuss the probe current in the next subsection.
The electric field pulse is induced at $t=0$ along $-y$ direction, and at $t=t_0$ along $+y$ direction for two dimensions.
Correspondingly the physical quantities are largely modified around these two characteristic time scales.
Since the results for the time range $0<t<t_0$ is the same as in the single-shot model, we focus on the regime $t>t_0$.
Figure~\ref{fig:3d_double_2nd_pump}(a) shows the time dependence of the superconducting fluctuation in the double pump-shot model.
The damping behavior is similar to that in the single pump-shot model: the power-law and exponential decay are observed.
At short time, however, there is almost no time dependence.
This is due to the remaining non-equilibrium contribution at $t=t_0$ induced from the first pump shot at $t=0$, which makes the difference from the single pump-shot model.
On the other hand, with strong pump, the power-law/logarithmic behavior is observed, which corresponds to the regime I in Fig.~\ref{fig:phase}.
Namely, while the regime II in the single pump-shot model may not be seen depending on the excitation process, the regime I is more robustly present for the case with a strong excitation pulse.

Figure~\ref{fig:3d_double_2nd_pump}(b) shows the electric current density.
In contrast to Fig.~\ref{fig:3d_double_2nd_pump}(a), the qualitative behaviors are the same as those in the single pump shot model even for small $t-t_0$ [see Fig.~\ref{fig:single_fluc_curr}(e)].
This is because the current induced from the first shot at $t=0$ is a simple decreasing function and almost vanishes at $t=t_0$ where the second pulse is induced.
Thus the characteristic power-law/logarithmic behaviors are not specific to the single pump-shot model.
We also show in Fig.~\ref{fig:3d_double_2nd_pump}(c) the time evolution of the free energies, which is again similar to the single pump-shot model.

\subsection{Paraconductivity in photo-excited systems}

In this subsection, we consider the probe current in the photo-excited state.
With the two dimensional double pump-shot model, the probe electric field is written as $\bm E_1(t) = \bm E_1 \cos \Omega (t-t_1)$,
and the current density in Eq.~\eqref{eq:j1} as
\begin{align}
	j_{1\mu}(t, \Omega) &= \sum_\nu \Big[
		\sg^{(1)}_{\mu\nu}(t, \Omega) \big( \cos \Omega (t-t_1) - 1 \big) \nonumber \\
		&\hspace{-.4mm}+ \sg^{(2)}_{\mu\nu}(t, \Omega) \sin \Omega (t-t_1)
		+ \sg^{(3)}_{\mu\nu}(t, \Omega)
	\Big]E_{1\nu},
\label{eq:paracond}
\end{align}
where $\mu,\nu=x,y,z$.
In equilibrium, if we take the time average, $\sg^{(1)}$ and $\sg^{(2)}$ represent the real and imaginary parts of the AC paraconductivity summarized in Appendix~\ref{app_eq}, and $\sg^{(3)}=0$.

\begin{figure}[t]
	\centering
	\includegraphics[width=85mm]{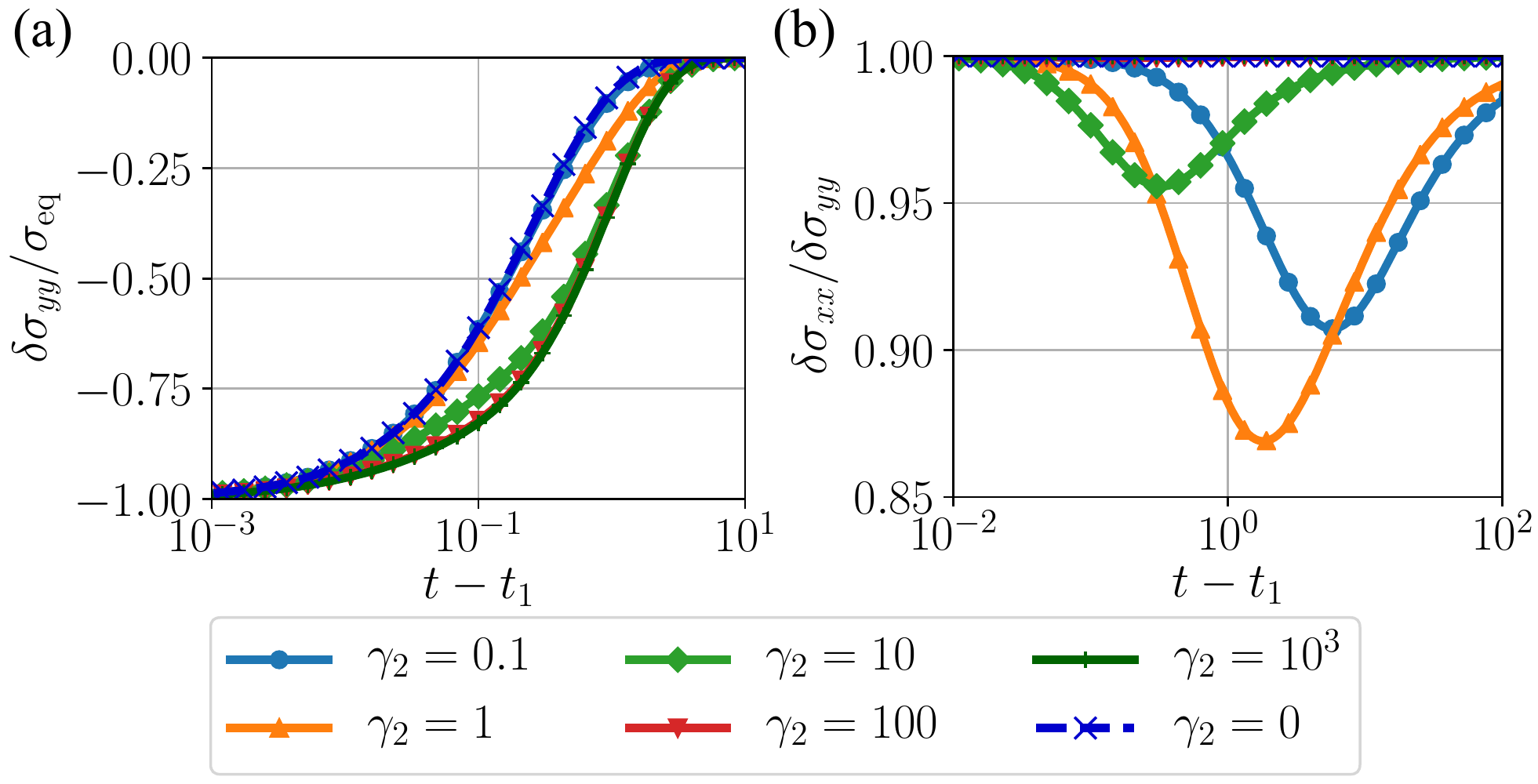}
	\caption{(a) Time dependence of the non-equilibrium component of the DC paraconductivity, which is normalized by its equilibrium component.
	(b) Anisotropy parameter $\delta \sg_{xx}/\delta \sg_{yy}$.
	Note that the vertical axis in (b) is measured from the unity(isotropic limit).
	We have taken $\gm_1=1$ and $t_1=1.1$ with the time unit $t_0$($=1$).
	}
	\label{fig:dc_paracond}
\end{figure}

\begin{figure}[t]
	\centering
	\includegraphics[width=85mm]{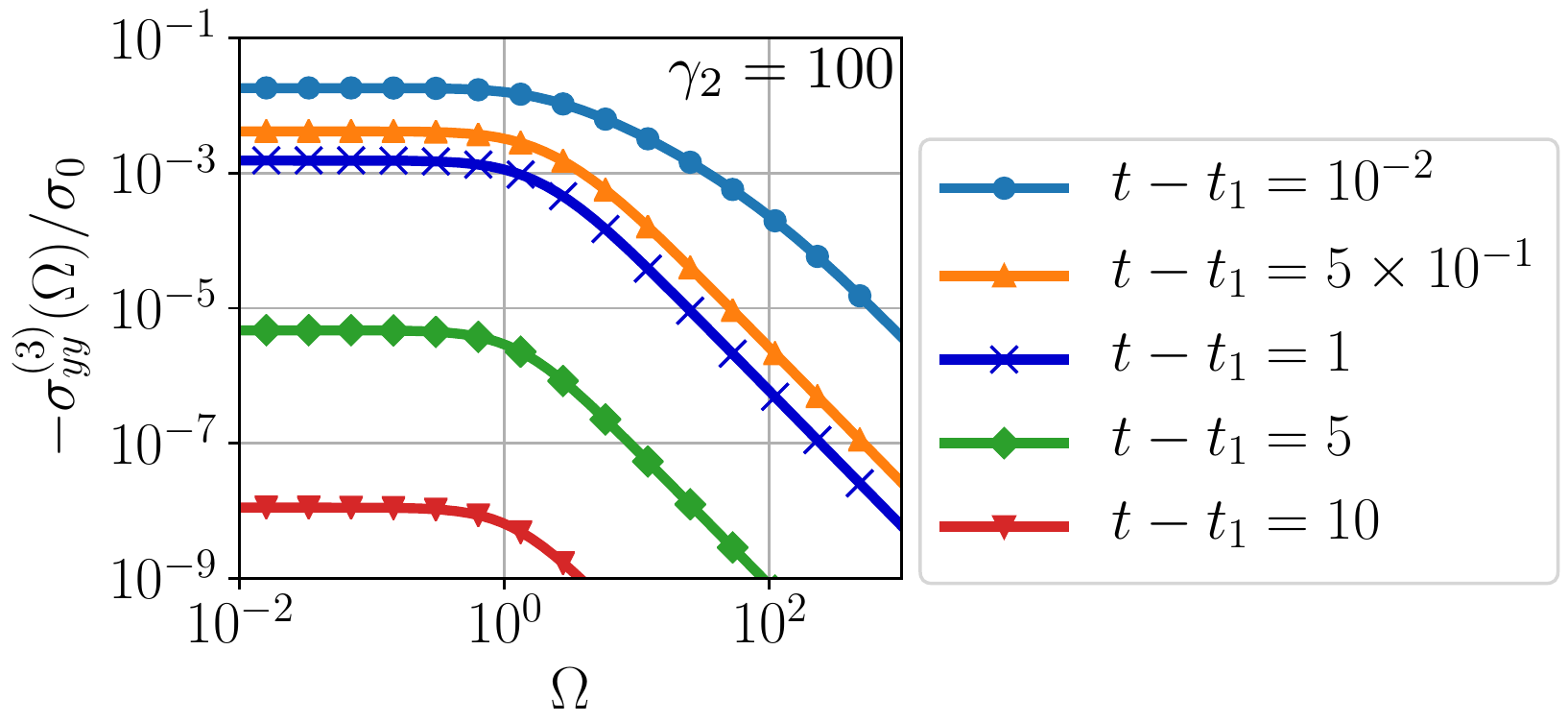}
	\caption{
		Frequency dependence of the AC paraconductivity $\sg^{(3)}$ at several different times.
		In these figures, we have taken $\gm_1 = 1$, $\gm_2=100$ and $t_1 = 1.1$ ($t_0=1$ is the unit of time).
		The normalization constant is defined by $\sg_0 = \frac{8e^2 C}{t_0\Gamma V}$.
	}
	\label{fig:ac_paracond}
\end{figure}

Let us first focus on the DC paraconductivity with $\Omega \to 0$.
In this case, the probe current measured from the equilibrium value is simply written as
\begin{align}
\delta j_{1\mu}(t) &= \delta \sg_{\mu\mu} (t) E_{1\mu},
\end{align}
which describes the characteristics of the non-equilibrium state.
Figure~\ref{fig:dc_paracond}(a) shows the DC paraconductivity in the photo excited states for $\gm_1 = 1$ and $t_1=1.1$, which is normalized by the equilibrium component of the conductivity ($\sg = \sg_{\rm eq} + \delta \sg$).
We note that $\delta \sg_{\mu\mu} (t)$ can be finite even for the case without pump (i.e. $\bm A_0=\bm 0$), since the sudden switching-on of the probing field turns the system into non-equilibrium.
As shown in Fig.~\ref{fig:dc_paracond}, the paraconductivity is decreased (i.e. $\delta \sg<0$) when the pump light $\bm A_0$ is induced.
This observation is consistent with the decrease of the superconducting fluctuation in Fig.~\ref{fig:3d_double_2nd_pump}(a).
Hence, this result cannot account for the experiments where the conductivity is largely enhanced \cite{Fausti11, Mitrano16}.
A relatively small change in Fig.~\ref{fig:dc_paracond} as a function $\gm_2$ is due to the choice of measuring time $t_1=1.1$: at this time the induced current has become a small value.
Since $\delta \sigma$ is the same order of magnitude as the equilibrium component $\sg_{\rm eq}$, the paraconductivity characteristic for non-equilibrium states can in principle be observed for the two and one dimensional systems where the paraconductivity $\sg_{\rm eq}$ is detectable.

With the pump process, the system becomes anisotropic because of the one-direction-oriented electric fields.
We show in Fig.~\ref{fig:dc_paracond}(b) the anisotropy parameter $\delta \sg_{xx}/ \delta \sg_{yy}$ for the non-equilibrium component of the paraconductivity.
This quantity shows a peaked structure as a function of time.
Namely, the photo-excited current makes the system anisotropic and for $t\to \infty$ the system relaxes into the isotropic equilibrium state.
The parameter $\gm_2$-dependence is also not simple.
For a weak pump such as $\gm_2=0.1$, the photo-induced current $\bm j_0$ is small and then the anisotropy is also small.
With increasing $\gm_2$, the anisotropy parameter increases with increasing photo-induced current.
However, for a sufficiently strong pump, the anisotropy becomes smaller again as shown in Fig.~\ref{fig:dc_paracond}(b).
This is because  the strong pump creates a quickly damping current with a large exponent as shown in Eq.~\eqref{eq:exponent_curr_1}, and little current is left when probing current is induced.
Thus the anisotropy behaves in a non-trivial manner as a function of $\gm_2$.

Finally we briefly discuss the AC paraconductivity in the photo-excited state.
We can separate the time and frequency dependences of $\delta \sg^{(1,2)}$, which are written as
\begin{align} \label{def_al}
	\delta \sg^{(1)}(t, \Omega)
	=&
 	\frac{8e^2}{m\Gamma V} \frac{\delta \mathcal{F}_{\mathrm{kin}}^{(\mu)} (t)}{\Omega^2} ,
	\\
	\delta \sg^{(2)}(t, \Omega)
	=&
	\frac{2e^2}{m} \frac{\delta \la |\psi (\bm{r}, t)|^2 \ra}{\Omega}  , \label{def_beta}
\end{align}
where $\mu$ means the direction where the pump pulse is shotted.
These forms can be understood from Eqs.~\eqref{def_current}, \eqref{eq:j0}, \eqref{eq:j1}, \eqref{double_fluc} and \eqref{eq:paracond}.
The time dependence of these quantities have already discussed in the above and new plots are not necessary.

The behaviors of $\sg^{(3)}$ are more complex because we cannot separate the time $t$ and frequency $\Omega$ dependences.
Figure~\ref{fig:ac_paracond} shows the frequency dependence of $\sg^{(3)}(t, \Omega)$ at different times.
We can see from Fig.~\ref{fig:ac_paracond} that $\sg^{(3)}$ is also a simple decreasing function of both time and frequency.
Since $\sg^{(3)}$ behaves as a constant for $\Omega \to 0$ and $\sim \Omega^{-2}$ for the large $\Omega$ limit, its shape is Lorentzian-like.
Thus all the frequency dependences are characterized by the three components $\delta \sg^{(1,2,3)}(t,\Omega)$.

\section{Summary} \label{sec_summ}

Based on the time-dependent Ginzburg-Landau (TDGL) theory, we have clarified the properties of superconducting fluctuation for the systems with the photo-excitation.
While the long-time behavior approaching to equilibrium is characterized by the exponential decay with a time constant determined by the temperature, in the intermediate regime the non-trivial power-law/logarithmically decaying regimes are identified.
The power-law/logarithmic decay is usually observed in the systems located at the critical point, but it can be seen in the photo-excited system even away from the critical point and the exponents are different depending on the excitation strength.
This is interpreted as the extension of the critical point into a finite regime by the photo-excitation.
Although the superconducting fluctuation is divergent at the critical point in the equilibrium case, in the excited state it is reduced from the value in equilibrium for the whole time range.

In addition to the properties of excited states, we have also formulated and calculated the probing currents using the linear-response theory.
We have measured the anisotropy of the conductivity in the excited state, which becomes maximal at the intermediate photo-excitation strength, in contrast to a naive expectation that the stronger pump just creates a larger anisotropic state.
This behavior is closely related to the exponent change in the current density as a function of the excitation amplitude in the power-law decaying regime.
We have also clarified the time evolution of the AC conductivities.
The probing currents discussed in this paper may be experimentally observed in one and two-dimensional systems since the magnitude of the non-equilibrium components of the linear-response conductivity is comparable to the equilibrium one.

These results are obtained based on the simple TDGL equation, which is applicable to a wide class of materials, and gives a foundation for further exploring the fluctuation phenomena in non-equilibrium superconductors.

\section*{Acknowledgement}

S.H. acknowledges N. Nagaosa for valuable discussions on paraconductivity.
The authors thank P. Werner for useful comments on our work and H. Shinaoka for his advise on the Gauss-Legendre method.
This work was supported by Japan Society for Promotion of Science (JSPS) KAKENHI Grants No.~JP16K17729 (N.T.), No.~JP18K13490 and No.~JP18H01176 (S.H.).

\appendix

\section{Dimensionless forms of physical quantities} \label{app_dimless}

\subsection{Definition of dimensionless functions}

In this section, we define dimensionless functions to evaluate the physical quantities numerically.
First of all, we introduce the unit time $\mathcal{T}$ which is chosen as $\gm_1^{-1}$ for the single pump-shot model and as $t_0$ for the double pump-shot model, to make the notation simple.
In addition, we define the constants as
\begin{align}
\hspace{-1mm}
q_0
=
\sqrt{\frac{2m\Gamma}{\mathcal{T}}}
,\ \
&k
=
\gm_1 \mathcal{T}
,\ \
\bm{x}_0
=
\sgn (e) \sqrt{\gm_2 \mathcal{T}} \frac{\bm{A}_0}{|\bm{A}_0|},
\end{align}
where the definition of $\gm_1$ and $\gm_2$ are given in Eqs.~\eqref{def_gm1} and \eqref{def_gm2}.
We have chosen $x_0 > 0$ in this paper.
Using these constants, the wavevector is written as $\bm{q} - 2e\bm{A}_0 = q_0 (\bm{x} - \bm{x}_0)$.
We define the several dimensionless functions in $d$ dimensional system as
\begin{align}
	X(a,b)
	&=
	\int \frac{\diff\bm x}{(2\pi)^d} \frac{1}{\bm x^2+k}
	\epn^{-(a+b)\left( \bm{x} - \frac{b}{a+b}\bm{x}_0 \right)^2},
	\\
	\tilde X(a,b)
	&=
	\int \frac{\diff\bm x}{(2\pi)^d} \frac{1}{(\bm{x} - \bm{x}_0)^2+k}
	\epn^{-(a+b)\left( \bm{x} - \frac{b}{a+b}\bm{x}_0 \right)^2},
	\\
	Y_\mu(a,b)
	&=
	\int \frac{\diff\bm x}{(2\pi)^d} \frac{x_\mu}{\bm x^2+k}
	\epn^{-(a+b)\left( \bm{x} - \frac{b}{a+b}\bm{x}_0 \right)^2},
	\\
	\tilde Y_\mu(a,b)
	&=
	\int \frac{\diff\bm x}{(2\pi)^d} \frac{x_\mu}{(\bm{x} - \bm{x}_0)^2+k}
	\epn^{-(a+b)\left( \bm{x} - \frac{b}{a+b}\bm{x}_0 \right)^2},
	\\
	Z_{\mu\nu}(a,b)
	&=
	\int \frac{\diff\bm x}{(2\pi)^d} \frac{x_\mu x_\nu}{\bm x^2+k}
	\epn^{-(a+b)\left( \bm{x} - \frac{b}{a+b}\bm{x}_0 \right)^2},
	\\
	\tilde Z_{\mu\nu}(a,b)
	&=
	\int \frac{\diff\bm x}{(2\pi)^d} \frac{x_\mu x_\nu}{(\bm{x} - \bm{x}_0)^2+k}
	\epn^{-(a+b)\left( \bm{x} - \frac{b}{a+b}\bm{x}_0 \right)^2},
	\\
	W(a,c)
	&=
	\frac{1}{d}\int \frac{\diff\bm x}{(2\pi)^d} \frac{\bm x^2(\bm x^2+k)\epn^{-a\bm x^2}}{(\bm x^2 + k)^2+ c^2},
	\\
	&\xrightarrow[c \to 0]{}
	W(a, 0)
	-
	W^{(2)} (a) c^2 + \mathcal{O}\left(c^4\right),
	\\
	W^{(2)}(a)
	&=
	\frac{1}{d} \int \frac{\diff\bm x}{(2\pi)^d} \frac{\bm{x}^2\epn^{-a\bm{x}^2}}{\left(\bm{x}^2 + k\right)^3},
	\\
	V(a, c)
	&=
	-\frac{1}{d} \int \frac{\diff \bm x}{(2\pi)^d} \frac{\bm x^2 \epn^{-a\bm x^2}}{(\bm x^2 + k)^2 [(\bm x^2 + k)^2 + c^2]}.
\end{align}
which are evaluated for each dimension separately.
The limit $c \to 0$ corresponds to the limit $\Omega \to 0$.
The function $W^{(2)}$ is used when we consider the DC paraconductivity.
We note that $Y_{\mu}(a, 0) = 0$ since the integrand is odd function, and there is the relation
$Z_{\mu \nu}(a, 0) = \delta_{\mu \nu}W(a, 0)$.

\subsubsection{Three-dimensional system}
We can choose $\bm A_0 = A_0 \hat {\bm z}$ without loss of generality.
Then we have the relations
\begin{align}
&Y_x=Y_y=0,
\\
&Z_{\mu\nu} = 0 \ \ (\mu\neq\nu),
\\
&Z_{xx} = Z_{yy},
\end{align}
since the integrand is an odd function.
We obtain the detailed functional forms as
\begin{align}
X(a,b) &= \frac{1}{4\pi^2 \sqrt{a+b}} G_0 \left( k(a+b), \frac{2bx_0}{\sqrt{a+b}} \right),
\\
Y_z(a,b) &= \frac{1}{4\pi^2 (a+b)} G_1 \left( k(a+b), \frac{2bx_0}{\sqrt{a+b}} \right),
\\
Z_{zz}(a,b) &= \frac{1}{4\pi^2 (a+b)^{3/2}} G_2 \left( k(a+b), \frac{2bx_0}{\sqrt{a+b}} \right),
\\
Z_{xx}(a,b)
&=
\frac{\sqrt \pi}{16\pi^2(a+b)^{3/2}}
-\frac{k}{2} X(a,b)
-\frac{1}{2} Z_{zz}(a,b).
\end{align}
Here the function $G_k(a,b)$ is defined by
\begin{align}
G_k(a,b) &= \epn^{-\frac{b^2}{4}}\int_0^\infty \diff x \, \frac{x^{k+2} \epn^{- x^2} }{x^2 + a} \int_{-1}^{1} \diff s \,  s^k \epn^{b xs}
\\
&=
\epn^{-\frac{b^2}{4}}
\sum_{\sigma=\pm}\sum_{l=0}^{k}
\sigma^{l-1} b^{l-k-1} a_l^{(k)} F_{l+1}(a,\sigma b),
\end{align}
where
\begin{align}
\{ a_{l}^{(0)} \} &= 1,
\\
\{ a_{l}^{(1)} \} &= -1, 1,
\\
\{ a_{l}^{(2)} \} &= 2, -2, 1,
\\
\{ a_{l}^{(3)} \} &= -6, 6, -3, 1,
\end{align}
and
\begin{align} \label{def_func_F}
F_n(a,b)
=
\int_0^\infty \diff x \, \frac{x^{n} \epn^{- (x -\frac{b}{2})^2} }{x^2 + a}.
\end{align}

\subsubsection{Two-dimensional system}
We can choose $\bm A_0 = A_0 \hat {\bm y}$ without loss of generality.
Then, we
use the relations
\begin{align}
    Y_x &= 0,
    \\
    Z_{\mu\nu} &= 0 \ \ (\mu \neq \nu).
\end{align}
We get
\begin{align}
    X(a,b) &= \frac{1}{2\pi} G^{(1)} \left( k(a+b), \frac{2bx_0}{\sqrt{a+b}} \right),
    \\
    Y_y(a,b) &= \frac{1}{2\pi \sqrt{a+b}} G^{(2)} \left( k(a+b), \frac{2bx_0}{\sqrt{a+b}} \right),
    \\
    Z_{yy}(a,b) &=
    \frac{1}{4\pi (a+b)} G^{(3)} \left( k(a+b), \frac{2bx_0}{\sqrt{a+b}} \right),
    \\
    Z_{xx}(a,b) &=
    \frac{1}{4\pi (a+b)} G^{(4)} \left( k(a+b), \frac{2bx_0}{\sqrt{a+b}} \right).
\end{align}
We have defined the new functions
\begin{align}
    G^{(1)} (a, b) =& \int_0^\infty \diff x\, \frac{x\epn^{-(x - \frac{b}{2})^2}}{x^2 + a} \tilde{I}_0(bx),
    \\
    G^{(2)} (a, b) =& \int_0^\infty \diff x\, \frac{x^2\epn^{-(x - \frac{b}{2})^2}}{x^2 + a} \tilde{I}_1(bx),
    \\
    G^{(3)} (a, b) =& \int_0^\infty \diff x\, \frac{x^3\epn^{-(x - \frac{b}{2})^2}}{x^2 + a} [\tilde{I}_0(bx) + \tilde{I}_2(bx)],
    \\
    G^{(4)} (a, b) =& \int_0^\infty \diff x\, \frac{x^3\epn^{-(x - \frac{b}{2})^2}}{x^2 + a} [\tilde{I}_0(bx) - \tilde{I}_2(bx)],
\end{align}
where $\tilde{I}_n(z) = \epn^{-|z|} I_n(z)$ and $I_n(z)$ is a modified Bessel function of the first kind.

\subsubsection{One-dimensional system}
We obtain the dimensionless functions in the following forms:
\begin{align}
    X(a,b) &= \frac{\sqrt{a+b}}{2\pi} \sum_{s=\pm} F_0 \left( k(a+b), \frac{2bx_0s}{\sqrt{a+b}} \right),
    \\
    Y_x(a,b) &= \frac{1}{2\pi} \sum_{s=\pm} sF_1 \left( k(a+b), \frac{2bx_0s}{\sqrt{a+b}} \right),
    \\
    Z_{xx}(a,b) &= \frac{1}{2\pi \sqrt{a+b}} \sum_{s=\pm} F_2 \left( k(a+b), \frac{2bx_0s}{\sqrt{a+b}} \right),
\end{align}
where the function $F_n(a, b)$ is given in Eq.~\eqref{def_func_F}.

\subsection{Single pump-shot model}

We consider the physical quantities in the single pump-shot model.
We first define the constants by
\begin{align}
	J_0 =& \frac{eC\mathcal{T}q_0^{d+1}}{m\Gamma^2V},
	\\
	\sg_0 =& \frac{2e^2C\mathcal{T}^3q_0^{d+2}}{m^2\Gamma^3V},
	\\
	\psi_0^2 =& \frac{C\mathcal{T}q_0^d}{\Gamma^2V},
	\\
	\mathcal{F}_0 =& \frac{C\mathcal{T}q_0^{d+2}}{4m\Gamma^2}.
\end{align}
The electric current density defined in Eqs.~\eqref{def_current}, \eqref{eq:j0} and \eqref{eq:j1} is rewritten as
\begin{widetext}
	\begin{align}
		\delta j_\mu (t, \Omega)
		=
		&\epn^{-kt/\mathcal{T}}
		\Bigg[
		-J_0 \tilde Y_\mu \left(\frac{t}{\mathcal{T}}, 0\right)
		+ \sg_0\frac{E_{1\mu}^s(t)}{2\Omega \mathcal{T}} X\left(0, \frac{t}{\mathcal{T}}\right)
		+ \sg_0
		\sum_\nu \frac{E_{1\nu}^c (t)}{\Omega^2 \mathcal{T}^2} \tilde Z_{\mu\nu} \left(\frac{t}{\mathcal{T}}, 0 \right)
		\Bigg]
		\nonumber \\
		&-
		\epn^{-kt/\mathcal{T}}
		\Bigg[
		\sg_0\frac{E_{1\mu}^s(t)}{2\Omega \mathcal{T}} X\left(\frac{t}{\mathcal{T}},0 \right)
		+ \sg_0
		\frac{E_{1\mu}^c (t)}{\Omega^2 \mathcal{T}^2} W\left(\frac{t}{\mathcal{T}}, 0 \right)
		\Bigg]
		+
		\epn^{-k(t-t_1)/\mathcal{T}}
		\sg_0
		E_{1\mu}
		V\left(\frac{t-t_1}{\mathcal{T}}, \Omega \mathcal{T} \right),
	\end{align}
\end{widetext}
where we have defined $\bm E_1^s (t) = \bm E_1 \sin \Omega (t - t_1)$ and $\bm E_1^c (t) = \bm E_1 \left[ \cos \Omega (t - t_1) -1\right]$ to make the notations simple.
When we take $\Omega \to 0$ limit, we expand these expressions up to the second order.

The deviation of the superconducting fluctuation originating from the pump is given by
\begin{align} \label{num_single_fluc}
    &\frac{\delta \la |\psi(\bm{r}, t)|^2 \ra}{\psi_0^2}
    =
	\epn^{-kt/\mathcal{T}} X\left(0, \frac{t}{\mathcal{T}}\right)
    -\epn^{-kt/\mathcal{T}} \tilde{X}\left(0, \frac{t}{\mathcal{T}}\right).
\end{align}
We also consider the GL free energy.
As shown in Eq.~\eqref{def_free_energy_component}, the fluctuation contribution $\delta \mathcal{F}_{\mathrm{fluct}}(t)$ is written as
\begin{align} \label{def_num_free_fluc}
    \frac{\delta \mathcal{F}_{\mathrm{fluct}}(t)}{\mathcal{F}_0}
    =
    k\,\frac{\delta \la |\psi(\bm{r}, t)|^2 \ra }{\psi_0^2} ,
\end{align}
where the superconducting fluctuation is given in Eq.~\eqref{num_single_fluc}.
The kinetic component of the GL free energy is written as
\begin{align}
	\frac{\delta \mathcal{F}_{\mathrm{kin}}^{(\mu)} (t)}{\mathcal{F}_0}
	&=
	\epn^{-kt/\mathcal{T}} Z_{\mu \mu}\left( 0, \frac{t}{\mathcal{T}} \right)
	-2 x_{0\mu} \epn^{-kt/\mathcal{T}} Y_{\mu}\left( 0, \frac{t}{\mathcal{T}} \right)
	\nonumber
	\\
	&\hspace{-6mm}
	+x_{0\mu}^2 \epn^{-kt/\mathcal{T}} X\left( 0, \frac{t}{\mathcal{T}} \right)
	-\epn^{-kt/\mathcal{T}} W \left( \frac{t}{\mathcal{T}}, 0 \right). \label{num_single_free_kin}
\end{align}

\subsection{Double pump-shot model}

Next, we consider the double pump-shot model.
The electric current density $\delta j_{\mu}(t, \Omega)$ is rewritten as
\begin{widetext}
	\begin{align}
	&\delta j_\mu (t, \Omega) = \epn^{-kt/\mathcal{T} -\frac{(t-t_0)t_0}{t\mathcal{T}}x_0^2}
	\Bigg[
	J_0 Y_\mu \left(\frac{t-t_0}{\mathcal{T}}, \frac{t_0}{\mathcal{T}}\right)
	+ \sg_0\frac{E_{1\mu}^s(t)}{2\Omega \mathcal{T}} X\left(\frac{t-t_0}{\mathcal{T}}, \frac{t_0}{\mathcal{T}}\right)
	+ \sg_0
	\sum_\nu \frac{E_{1\nu}^c (t)}{\Omega^2 \mathcal{T}^2} Z_{\mu\nu} \left(\frac{t-t_0}{\mathcal{T}}, \frac{t_0}{\mathcal{T}} \right)
	\Bigg]
	\nonumber \\
	&+
	\epn^{-k(t-t_0)/\mathcal{T}}
	\Bigg[
	J_0 \tilde Y_\mu \left(\frac{t-t_0}{\mathcal{T}}, 0\right)
	+ \sg_0\frac{E_{1\mu}^s(t)}{2\Omega \mathcal{T}} \tilde X\left(\frac{t-t_0}{\mathcal{T}}, 0\right)
	+ \sg_0
	\sum_\nu \frac{E_{1\nu}^c (t)}{\Omega^2 \mathcal{T}^2} \tilde Z_{\mu\nu} \left(\frac{t-t_0}{\mathcal{T}}, 0 \right)
	\Bigg]
	\nonumber \\
	&-
	\epn^{-kt/\mathcal{T} -\frac{(t-t_0)t_0}{t\mathcal{T}}x_0^2}
	\Bigg[
	J_0 \tilde Y_\mu \left(\frac{t-t_0}{\mathcal{T}}, \frac{t_0}{\mathcal{T}}\right)
	+ \sg_0 \frac{E_{1\mu}^s(t)}{2\Omega \mathcal{T}} \tilde X\left(\frac{t-t_0}{\mathcal{T}}, \frac{t_0}{\mathcal{T}}\right)
	+ \sg_0
	\sum_\nu \frac{E_{1\nu}^c (t)}{\Omega^2 \mathcal{T}^2} \tilde Z_{\mu\nu} \left(\frac{t-t_0}{\mathcal{T}}, \frac{t_0}{\mathcal{T}} \right)
	\Bigg]
	\nonumber \\
	&-
	\epn^{-k(t-t_0)/\mathcal{T}}
	\Bigg[
	\sg_0 \frac{E_{1\mu}^s(t)}{2\Omega \mathcal{T}} X\left(\frac{t-t_0}{\mathcal{T}},0 \right)
	+ \sg_0
	\frac{E_{1\mu}^c (t)}{\Omega^2 \mathcal{T}^2} W\left(\frac{t-t_0}{\mathcal{T}}, 0 \right)
	\Bigg]
	+
	\epn^{-k(t-t_1)/\mathcal{T}}
	\sg_0 E_{1\mu}
	V\left(\frac{t-t_1}{\mathcal{T}}, \Omega \mathcal{T} \right).
	\end{align}
\end{widetext}
The case with $0< t \leq t_0$ is the same situation corresponding to the single pump-shot model, which we have already considered in the above subsection.
We then consider the region $t > t_0$. The fluctuation is written as
\begin{align} \label{num_double_after_fluc}
    \frac{\delta \la |\psi (\bm{r}, t)|^2 \ra}{\psi_0^2}
    =&
    \epn^{-kt/\mathcal{T} - \frac{(t-t_0)t_0}{t\mathcal{T}}x_0^2} X\left(\frac{t-t_0}{\mathcal{T}}, \frac{t_0}{\mathcal{T}}\right)
	\nonumber
	\\
	&+\epn^{-k(t-t_0)/\mathcal{T}} \tilde{X}\left(\frac{t-t_0}{\mathcal{T}}, 0\right)
	\nonumber
	\\
	&-\epn^{-kt/\mathcal{T} - \frac{(t-t_0)t_0}{t\mathcal{T}}x_0^2} \tilde{X}\left(\frac{t-t_0}{\mathcal{T}}, \frac{t_0}{\mathcal{T}}\right)
	\nonumber
	\\
	&-\epn^{-k(t-t_0)/\mathcal{T}} X\left(\frac{t-t_0}{\mathcal{T}}, 0\right).
\end{align}
Next, we consider the GL free energy.
The fluctuation contribution is same as Eq.~\eqref{def_num_free_fluc}.
The kinetic component is
	\begin{align}
	    \frac{\delta \mathcal{F}_{\mathrm{kin}}^{(\mu)} (t)}{\mathcal{F}_0}
	    =&
		\epn^{-kt/\mathcal{T} - \frac{(t-t_0)t_0}{t\mathcal{T}}x_0^2} Z_{\mu \mu}\left( \frac{t-t_0}{\mathcal{T}}, \frac{t_0}{\mathcal{T}} \right)
		\nonumber
		\\
	    &+\epn^{-k(t-t_0)/\mathcal{T}} \tilde{Z}_{\mu\mu}\left( \frac{t-t_0}{\mathcal{T}}, 0 \right)
		\nonumber
		\\
		&-\epn^{-kt/\mathcal{T} - \frac{(t-t_0)t_0}{t\mathcal{T}}x_0^2} \tilde{Z}_{\mu \mu}\left( \frac{t-t_0}{\mathcal{T}}, \frac{t_0}{\mathcal{T}} \right)
		\nonumber
		\\
		&-\epn^{-k(t-t_0)/\mathcal{T}} W \left( \frac{t-t_0}{\mathcal{T}}, 0 \right).
	\end{align}

\section{Numerical evaluation for the dimensionless functions} \label{app_num}

The dimensionless functions defined in Appendix~\ref{app_dimless} are written in the general following form
\begin{align}
&\int_0^\infty \diff x\, \mathscr F(x)\epn^{-\left(x - \frac{b}{2}\right)^2}
=
\int_{-\frac{b}{2}}^\infty \diff x \, \mathscr F\Big(x + \frac{b}{2}\Big) \epn^{-x^2}
\\
\simeq &
\frac{L_\mathrm{ max} - L_\mathrm{ min}}{2}
\int_{-1}^{1} \diff y \, \mathscr F\Big(\tfrac{L_\mathrm{ max}-L_\mathrm{ min}}{2}y + \tfrac{L_\mathrm{ max}+L_\mathrm{ min}}{2} + \tfrac{b}{2}\Big)
\nonumber
\\
&\hspace{22mm}\times
\epn^{- \left(\tfrac{L_\mathrm{ max}-L_\mathrm{ min}}{2}y+\tfrac{L_\mathrm{ max}+L_\mathrm{ min}}{2}\right)^2},
\end{align}
where $L_\mathrm{max} > L_\mathrm{min}$ with the expressions
\begin{align}
\begin{matrix}
L_\mathrm{ min} = -L,\ \ L_\mathrm{ max} = L & (2L<b),
\\
L_\mathrm{ min} = -\frac{b}{2},\ \ L_\mathrm{ max} = 2L-\frac{b}{2} & (b<2L).
\end{matrix}
\end{align}
We assume that the function $\mathscr{F}(x)$ asymptotically behaves as a power function.
Then, due to the presence of the exponential function, it is safe to choose $L\simeq 5$, since $\epn^{-5^2}\simeq 1.4\times 10^{-11}$ are so small.
We have used the Gauss-Legendre method \cite{Hildebrand_book} when we integrate the function numerically.

\section{TDGL theory in equilibrium} \label{app_eq}

\begin{figure*}[t]
	\centering
	\includegraphics[width=160mm]{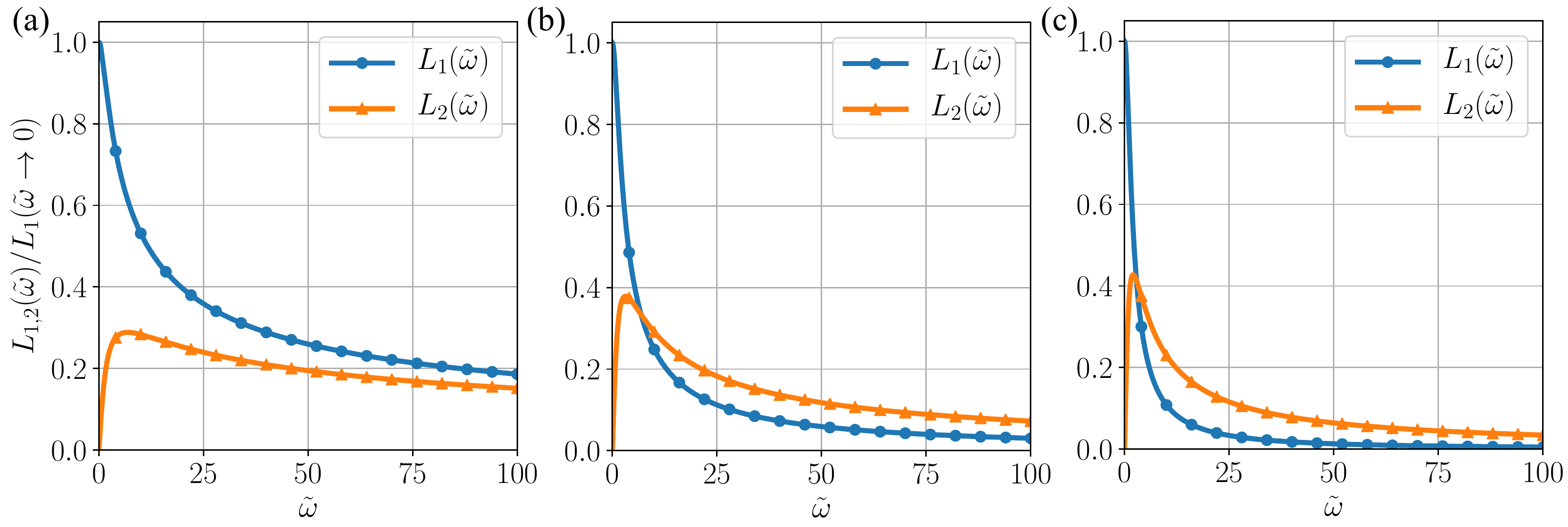}
	\caption{
		Frequency dependences of $L_1$ and $L_2$ in the (a) three-, (b) two- and (c) one-dimensional systems.
	}
	\label{fig:paracond_eq}
\end{figure*}

\subsection{General Formulation}

We review the TDGL theory in equilibrium system based on our formalism \cite{Larkin_book}.
We assume the weak vector potential $\bm{A}(t)$ without specifying its form.
The superconducting fluctuation with the wave vector is obtained as
	\begin{align}
		&\la |\psi_{\bm{q}}(t)|^2 \ra
		= \nonumber \\
		&\frac{C}{\Gamma^2 V} \int_{-\infty}^{t} \diff t^{\prime} \,
		\exp \left[
			-\frac{2}{\Gamma} \left( \frac{\bm{q}^2}{4m} + a\right) (t - t^{\prime})
		\right] \nonumber \\
		&\hspace{11mm} \times
        \left(
			1 + \frac{2e}{m\Gamma} \int_{t^{\prime}}^{t} \diff t^{\prime \prime} \, \bm{q} \cdot \bm{A} (t^{\prime \prime})
		\right)
		+\mathcal{O}\left( \bm{A}^2 \right),
	\end{align}
where we used the random average given by Eq.~\eqref{random_ave}.
We can expand the representation of the current up to the first order of $\bm{A}$ as
\begin{align}
	\bm{j}(t)
	&=
    \sum_{\bm{q}} \frac{e}{mV} \left[ \bm{q} - 2e\bm{A}(t) \right]
    \la |\psi_{\bm{q}}(t)|^2 \ra ,
    \\
    &\equiv
	\bm{j}_0(t)
	+\bm{j}_1(t)
	+\mathcal{O}\left( \bm{A}^2 \right),
\end{align}
where
\begin{align}
    \bm{j}_0(t)
    =&
    \frac{e}{mV} \sum_{\bm{q}} \bm{q} \la |\psi_{\bm{q}}(t)|^2 \ra_0 = 0,
    \\
    \bm{j}_1(t)
    =&
    \frac{e}{mV} \sum_{\bm{q}} \bm{q} \la |\psi_{\bm{q}}(t)|^2 \ra_1
    -\frac{2e^2}{mV} \bm{A}(t) \sum_{\bm{q}} \la |\psi_{\bm{q}}(t)|^2 \ra_0,
    \\
    \equiv &
    \bm{j}_1^P(t) + \bm{j}_1^D(t).
\end{align}
We have treated the vector potential as perturbation.
We note that the integrand of $\bm{j}_0(t)$ with respect to $\bm{q}$ is odd function.
The subscripts ``0" and ``1" mean the pertubation order of $\bm{A}$.
The symbols ``P" and ``D" mean ``Paramagnetic" and ``Diamagnetic" contributions.
We abbreviate the argument $\bm{r}$ since
the current is spatially uniform.

Performing the Fourier transformation,
the current is rewritten as
\begin{align}
	j_{\mu}(\omega)
	&=
    K_{\mu \nu}(\omega) A_{\nu}(\omega),
    \\
    &\equiv
	\left(
		K_{\mu\nu}^P(\omega) + K_{\mu\nu}^D(\omega)
	\right) A_{\nu}(\omega),
\end{align}
where
\begin{align}
	K_{\mu\nu}^P(\omega)
	=&
	\frac{2e^2C}{m^2\Gamma^3 V^2} \sum_{\bm{q}} q_{\mu} q_{\nu} \frac{g(\bm{q})^2}{1-\imu \omega g(\bm{q})},
	\\
	=&
	\frac{2e^2C}{m^2\Gamma^3 V^2} \sum_{\bm{q}} q_{\mu} q_{\nu}
	g(\bm{q})^2 \sum_{n=0}^{\infty} \left( \imu \omega g(\bm{q})\right)^n,
    \\
	K_{\mu\nu}^D(\omega)
	=&
	-\frac{2e^2C}{m\Gamma^2 V^2} \delta_{\mu\nu} \sum_{\bm{q}} g(\bm{q}).
\end{align}
The paraconductivity $\sigma_{\mu\nu}(\omega)$ is then obtained by
\begin{align}
	&\sigma_{\mu \nu}(\omega)
	=
	\frac{K_{\mu\nu}(\omega)}{\imu \omega}.
\end{align}
Note the relation $K_{\mu\nu}^P(\omega)|_{n=0} = -K_{\mu\nu}^D(\omega)$.
The total kernel can be rewritten as
\begin{align}
	K_{\mu\nu}(\omega)
	=
	\frac{2e^2C}{\imu m\Gamma^2 V^2 \omega} \delta_{\mu\nu} \sum_{\bm{q}} \sum_{n=1}^{\infty}
	\frac{\left( \imu \omega g(\bm{q}) \right)^{n+1}}{n+1}.
\end{align}
When we define a dimensionless frequency $\tilde{\omega} = \frac{\omega}{\gm_1} = \frac{\Gamma \omega}{2a}$, the response kernel is further rewritten as
\begin{align}
    K_{\mu\nu}(\omega)
    =
    \frac{e^2C}{\imu m\Gamma Va} \cdot \frac{1}{\tilde{\omega}} \delta_{\mu\nu} I(\tilde{\omega}),
\end{align}
where
\begin{align}
    I(\tilde{\omega}) = \int \frac{\diff \bm{q}}{(2\pi)^d} \left[
        -\imu \tilde{\omega} \frac{4ma}{\bm{q}^2 + 4ma}
        -\ln\left( 1-\imu \tilde{\omega} \frac{4ma}{\bm{q}^2 + 4ma}\right)
    \right].
\end{align}
In the following, we perform the $\bm{q}$-integral for each dimension.

\subsection{Three-dimensional system}

The AC paraconductivity for $d=3$ is given by
\begin{align}
	\sigma_{\mu\nu}(\omega)
	=
	\frac{e^2C\sqrt{m}}{3\pi V} a^{-1/2} \delta_{\mu\nu}
	\left[ L_1 \left( \frac{\Gamma \omega}{2a} \right) + \imu L_2 \left( \frac{\Gamma \omega}{2a} \right) \right],
\end{align}
where
\begin{align}
	&L_1(\tilde{\omega})
	= \nonumber \\
	&\frac{1}{\tilde{\omega}^2} \left[
		2 - \sqrt{2\left( \sqrt{1 + \tilde{\omega}^2} + 1\right)} + \tilde{\omega}\sqrt{2\left( \sqrt{1 + \tilde{\omega}^2} - 1\right)}
	\right],
	\\
	&L_2(\tilde{\omega})
	= \nonumber \\
	&\frac{1}{\tilde{\omega}^2} \left[
		-3\tilde{\omega} + \sqrt{2\left( \sqrt{1 + \tilde{\omega}^2} - 1\right)} + \tilde{\omega}\sqrt{2\left( \sqrt{1 + \tilde{\omega}^2} + 1\right)}
	\right].
\end{align}
The dependence of these quantities with respect to $\tilde{\omega}$ is shown in Fig.~\ref{fig:paracond_eq}(a).

\subsection{Two-dimensional system}

The AC paraconductivity for $d=2$ is given by
\begin{align}
	\sigma_{\mu\nu}(\omega)
	=
	\frac{e^2C}{2\pi V} a^{-1} \delta_{\mu\nu}
	\left[ L_1 \left( \frac{\Gamma \omega}{2a} \right) + \imu L_2 \left( \frac{\Gamma \omega}{2a} \right) \right] ,
\end{align}
where
\begin{align}
	L_1(\tilde{\omega})
	=
	&\frac{1}{\tilde{\omega}^2} \left[
		\tilde{\omega} \arctan \tilde{\omega} - \frac{1}{2} \ln \left( 1 + \tilde{\omega}^2\right)
	\right],
	\\
	L_2(\tilde{\omega})
	=
	&\frac{1}{\tilde{\omega}^2} \left[
		\arctan \tilde{\omega} + \frac{1}{2} \tilde{\omega} \ln \left( 1 + \tilde{\omega}^2 \right) - \tilde{\omega}
	\right].
\end{align}
These functions are shown in Fig.~\ref{fig:paracond_eq}(b).

\subsection{One-dimensional system}

The AC paraconductivity for $d=1$ is given by
\begin{align}
	\sigma_{\mu\nu}(\omega)
	=
	\frac{e^2C}{2V\sqrt{m}} a^{-3/2} \delta_{\mu\nu}
	\left[ L_1 \left( \frac{\Gamma \omega}{2a} \right) + \imu L_2 \left( \frac{\Gamma \omega}{2a} \right) \right] ,
\end{align}
where
\begin{align}
	L_1(\tilde{\omega})
	=
	&\frac{1}{\tilde{\omega}^2} \left[
		-2 + \sqrt{2\left( \sqrt{1 + \tilde{\omega}^2} + 1\right)}
	\right],
	\\
	L_2(\tilde{\omega})
	=
	&\frac{1}{\tilde{\omega}^2} \left[
		\tilde{\omega} - \sqrt{2\left( \sqrt{1 + \tilde{\omega}^2} - 1\right)}
	\right].
\end{align}
These functions are shown in Fig.~\ref{fig:paracond_eq}(c).

We note that the vertical axis in Fig.~\ref{fig:paracond_eq} is normalized by the $L_1(\tilde \omega \to 0)$.
For all the dimensions, $L_1$ ($L_2$), which is the real (imaginary) part of paraconductivity, is an even (odd) function of frequency.
These quantities satisfy the Kramers-Kronig relation.
As seen from Fig.~\ref{fig:paracond_eq}, the behavior in the low-frequency region is sharper for the lower dimensional system.

\section{Additional data for the double pump-shot model in three- and one-dimensional systems} \label{sec_app_add_fig}

\begin{figure}
	\includegraphics[width=85mm]{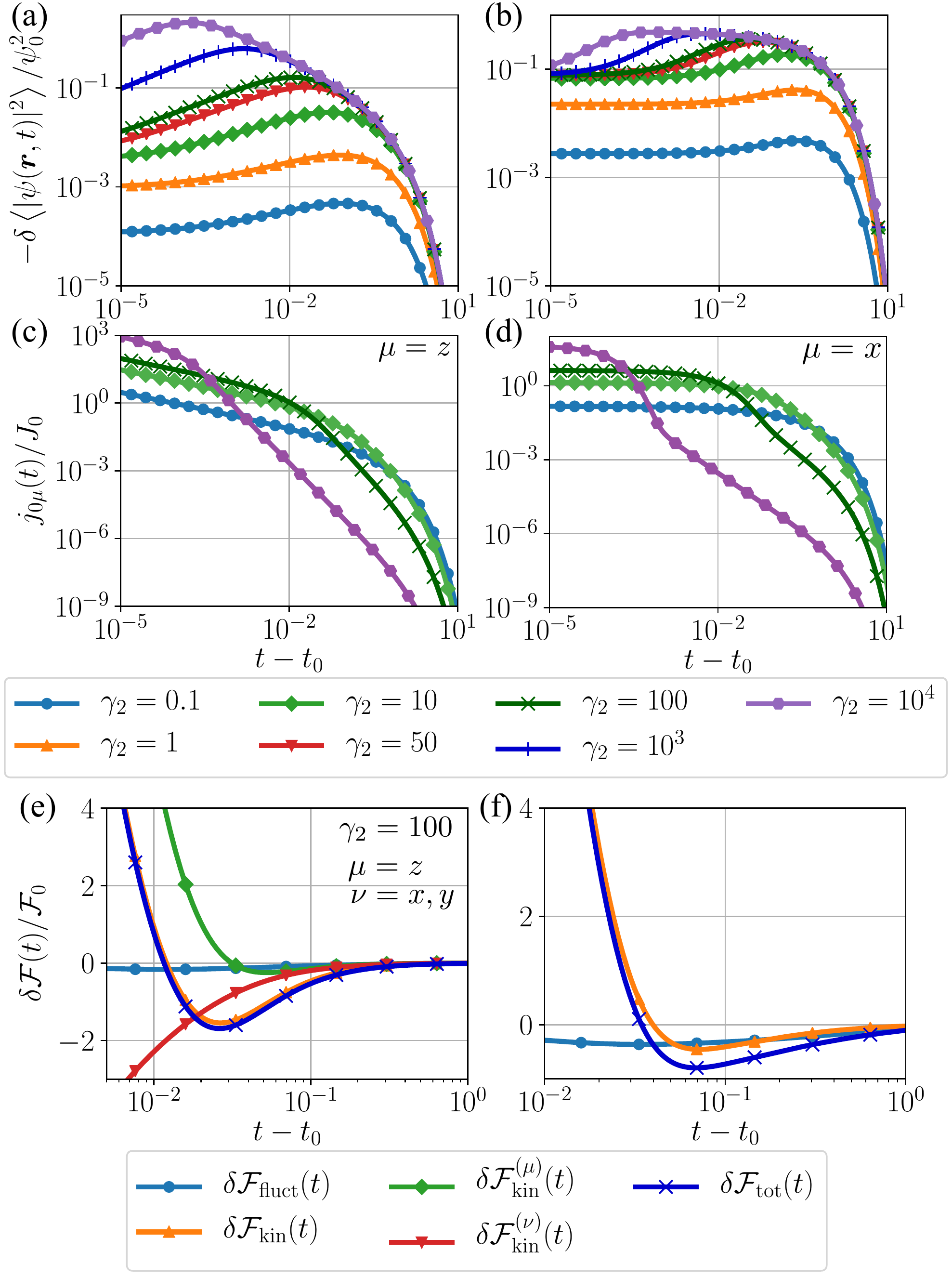}
	\caption{
		Time dependences of the (a,b) superconducting fluctuation, the (c,d) excited current and the (e,f) GL free energy with the logarithmic time scale in the double pump-shot model.
		The panels (a,c,e) are for three-dimensional systems and (b,d,f) for one-dimensional system.
		We take the parameter as $\gm_1 = 1$ and (e,f) $\gm_2 = 100$ to compare with the results of the single pump-shot model ($t_0 = 1$).
		$\mu$ and $\nu$ used in (e) have the same meanings as Fig.~\ref{fig:free_energy}(a).
		The normalized constants are defined by
		$\psi_0^2 = \frac{Ct_0}{\Gamma^2 V} \left( \frac{2m\Gamma}{t_0}\right)^{d/2}$,
		$J_0 = \frac{eCt_0}{m\Gamma^2 V} \left( \frac{2m\Gamma}{t_0}\right)^{\frac{d+1}{2}}$ and
		$\mathcal{F}_0 = \frac{C}{2\Gamma} \left( \frac{2m\Gamma}{t_0}\right)^{d/2}$.
	}
	\label{fig:1_3d_double_2nd_pump}
\end{figure}

\begin{figure}
	\includegraphics[width=85mm]{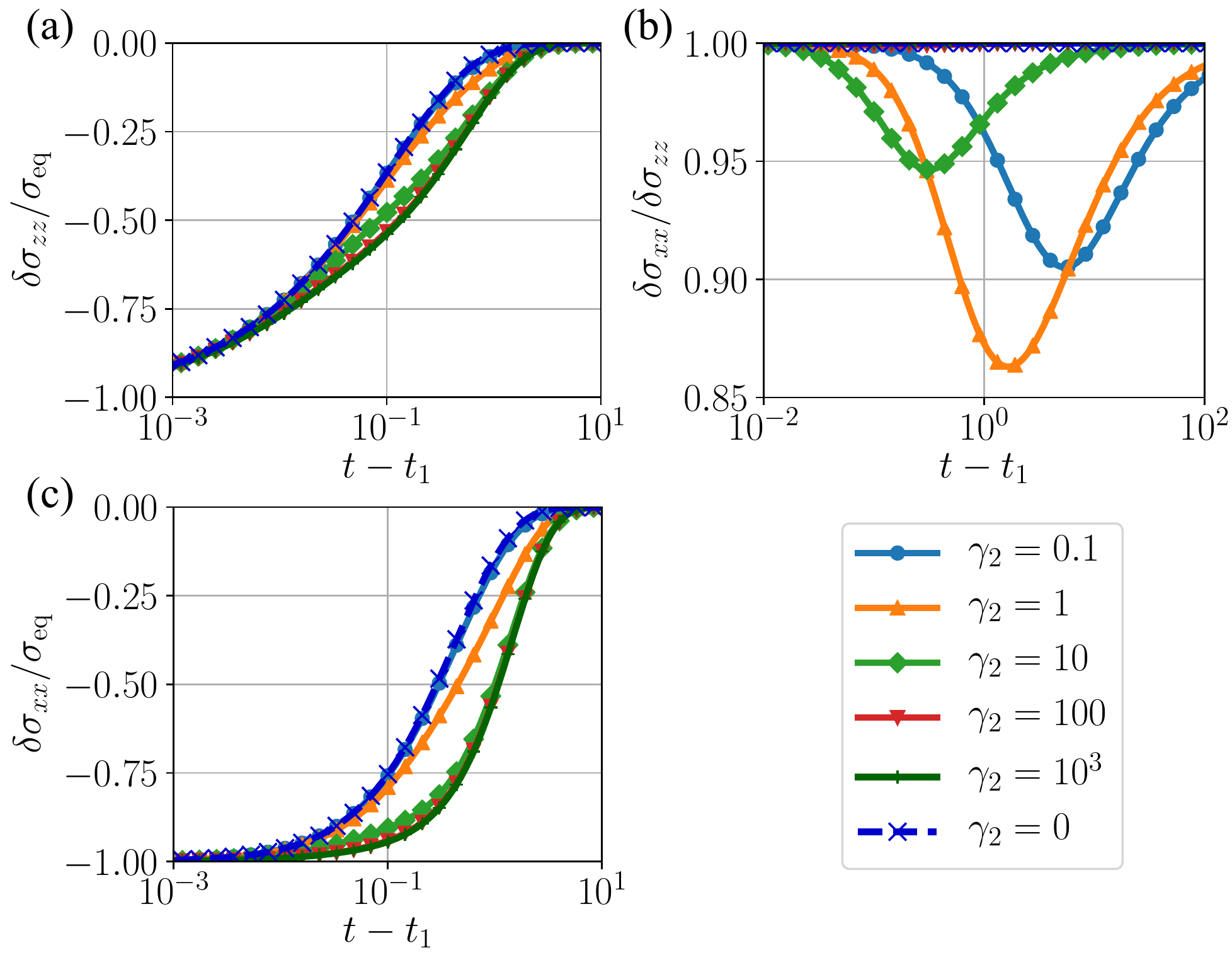}
	\caption{
		Time dependences of the non-equilibrium component of (a) the DC paraconductivity and (b) anisotropy parameter $\delta \sg_{xx}/\delta \sg_{zz}$ in the three-dimensional systems (double pump-shot model).
		Note that the vertical axis in (b) is measured from the unity (isotropic limit).
		The DC paraconductivity for one-dimensional systems is also shown in (c).
		We have taken $\gm_1=1$ and $t_1=1.1$ with the time unit $t_0$($=1$).
	}
	\label{fig:dc_paracond_1_3d}
\end{figure}

\begin{figure}
	\includegraphics[width=85mm]{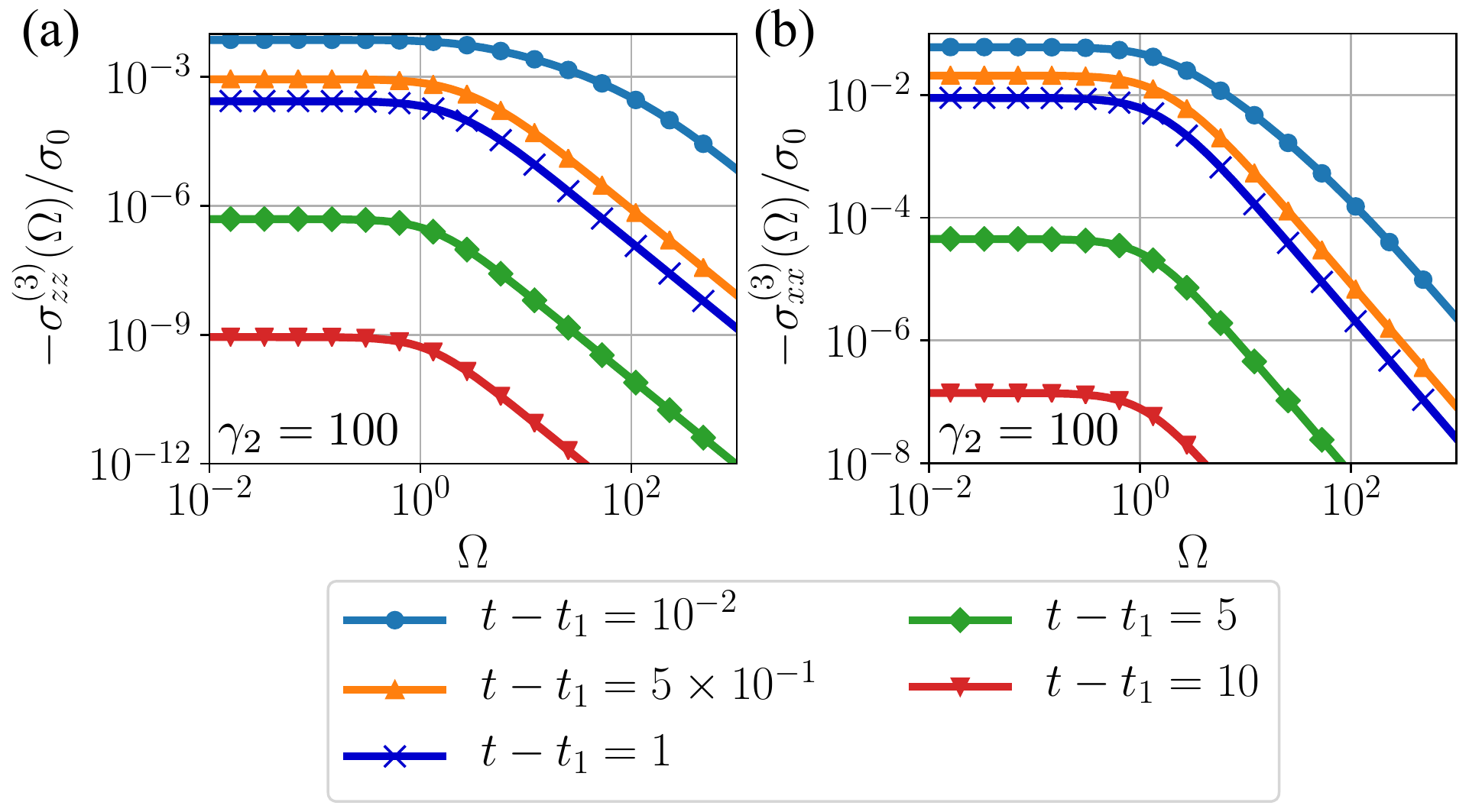}
	\caption{
		Frequency dependence of the AC paraconductivity $\sg^{(3)}$ at several different times in the (a) three- and (b) one-dimensional systems (double pump-shot model).
		In these figures, we have taken $\gm_1 = 1$, $\gm_2=100$ and $t_1 = 1.1$ with the unit time $t_0$ ($=1$).
		The normalization constant is defined by $\sg_0 = \frac{4e^2 C}{m\Gamma^2 V} \left( \frac{2m\Gamma}{t_0}\right)^{d/2}$.
	}
	\label{fig:ac_paracond_1_3d}
\end{figure}

While the double pump-shot model in the two-dimensional system is studied in the main text, here we show the results in three and one dimensions to make the data complete.
Figure~\ref{fig:1_3d_double_2nd_pump} shows the time dependence of the fluctuation, current density and free energy, which are the plots similar to Fig.~\ref{fig:3d_double_2nd_pump}.
Figures~\ref{fig:dc_paracond_1_3d} and \ref{fig:ac_paracond_1_3d} show the non-equilibrium component of the DC and AC paraconductivties, respectively, which correspond to Figs.~\ref{fig:dc_paracond} and \ref{fig:ac_paracond} of the main text.

\end{document}